\documentclass[floatfix,aps,twocolumn,preprintnumbers,amsmath,amssymb,nofootinbib,superscriptaddress,showkeys,showpacs]{revtex4}
\usepackage{graphicx,color}
\usepackage{amsmath,amssymb,bm}
\usepackage{dcolumn}

\begin{document}

\title{ Global analysis of the COVID-19 pandemic using simple
  epidemiological models }

\author{Jos\'e Enrique Amaro}\email{amaro@ugr.es}
\affiliation{
              Departamento de F\'isica At\'omica, Molecular y Nuclear and Instituto Carlos I de F\'isica Te\'orica y Computacional, 
              Universidad de Granada, E-18071 Granada, Spain.}

\author{J\'er\'emie Dudouet }             \email{j.dudouet@ip2i.in2p3.fr} 
\affiliation{ Univ Lyon, Univ Claude Bernard Lyon 1, CNRS/IN2P3, IP2I Lyon, UMR 5822, F-69622, Villeurbanne, France }

\author{Jos\'e Nicol\'as Orce}
             \email{jnorce@uwc.ac.za} 
\affiliation{Department of Physics \& Astronomy, University of the Western Cape, P/B X17 Bellville ZA-7535, South Africa.}

\date{\today}

\begin{abstract}

Several analytical models have been used in this work to describe the evolution of death cases arising from coronavirus ({\small COVID-19}). 
The Death or `D' model is a simplified version of the SIR (susceptible-infected-recovered) model,  
which assumes no recovery over time, and allows for the transmission-dynamics equations to be solved analytically.  
The D-model can be extended to describe various focuses of infection, which may account for the original pandemic (D1), the lockdown (D2) and 
other effects (Dn). 
The evolution of the {\small COVID-19} pandemic in several countries (China, Spain, Italy, France, UK, Iran, USA and Germany) 
shows a similar behavior in concord with the D-model trend, characterized by a rapid  increase of death cases followed by a slow decline, 
which are affected by the earliness and efficiency of the lockdown effect.
These results are in agreement with more accurate calculations using the extended SIR model 
with a parametrized solution and more sophisticated Monte Carlo grid simulations, which predict similar trends and 
indicate a common evolution of the pandemic with universal parameters.


\end{abstract}

\maketitle

\section{Motivation}
\label{intro}


The SIR (susceptible-infected-recovered) model 
is widely used as first-order approximation to viral spreading of contagious epidemics~\cite{sir}, 
mass immunization planning~\cite{vaccine,vaccine2}, marketing, informatics and social networks~\cite{2}. 
Its cornerstone is the so-called ``mass-action'' principle introduced by Hamer, which assumes that the course of an epidemic depends on the rate 
of contact between susceptible and infected individuals~\cite{hamer}. This idea was extended to a continuous time framework by Ross in 
his pioneering work on malaria transmission dynamics~\cite{ross,ross2,rosshudson}, and finally put into its classic mathematical form by Kermack 
and McKendric~\cite{KM}. The SIR model was further developed by Kendall, who provided a spatial generalization of the Kermack and  McKendrick model in a closed population~\cite{kendall1957} (i.e. neglecting the effects of spatial migration), and Bartlett, who -- after investigating the connection between the periodicity of measles epidemics  and community size -- predicted a traveling wave of infection moving out from the initial source of infection~\cite{bartlett1957,bartlett1957_2}. 
More recent implementations have considered the typical incubation period of the disease and the spatial migration of the
population.

The  {\small COVID-19} pandemic has ignited the submission of multiple  manuscripts in the last weeks. Most statistical distributions used to estimate disease occurrence  are of the binomial, Poisson, Gaussian, Fermi or exponential types. Despite their intrinsic differences, these distributions generally lead to 
similar results, assuming independence and homogeneity of disease risks~\cite{anydist}.

In this work, we propose a simple and easy-to-use epidemiological model -- the Death or D model~\cite{amaro} -- that can be compared with data in order to investigate the evolution of the infection and deviations from the predicted trends. The D model is a simplified version of the SIR model with analytical solutions under the assumption of no recovery -- at least during the time of the pandemic. We apply it globally to countries where the infestation of the {\small COVID-19} 
coronavirus has widespread and caused thousands of deaths~\cite{corona1,corona2}. 

Additionally, D-model calculations are benchmarked with more sophisticated and reliable calculations 
using the extended SIR (ESIR) and Monte Carlo Planck (MCP) models -- also developed in this work -- which provide similar results, but allow for a more coherent spatial-time disentanglement of the various effects present during a pandemic. 
A similar ESIR model has recently been proposed by Squillante and collaborators for infected individuals as a function of time, 
based on the Ising model -- which describes ferromagnetism in statistical mechanics -- 
and a Fermi-Dirac distribution~\cite{ising}. This model also reproduces \emph{a posteriori} the {\small COVID-19} data for infestations in China 
as well as other pandemics such as Ebola, SARS, and influenza A/H1N1.\\


The SIR model considers the three possible states of the members of a closed population affected by a contagious disease. 
It is, therefore, characterized by a system of three coupled non-linear ordinary differential equations~\cite{3}, which involve 
three time-dependent functions: 
\begin{itemize}
 \item Susceptible individuals, $S(t)$, at risk of becoming infected by the disease.
 \item Infected individuals, $I(t)$.
 \item Recovered or removed individuals, $R(t)$, who were infected and may have developed an immunity system or die.
\end{itemize}

The SIR model describes well a viral disease, where individuals typically go from the susceptible class $S$ to the infected class $I$, and finally to 
the removed class $R$. 
Recovered individuals cannot go back to be susceptible or infected classes, as it is, potentially, the case of bacterial infection. 
The resulting transmission-dynamics system for a closed population is described by

\begin{eqnarray}
 \frac{dS}{dt}&=&-\lambda SI, \label{eq:10}\\ 
\frac{dI}{dt}&=& \lambda SI - \beta I,  \label{eq:11} \\ 
 \frac{dR}{dt}&=& \beta I,  \label{eq:12} \\ 
 N&=&S(t)+I(t)+R(t),  \label{eq:13}
\end{eqnarray}
where $\lambda > 0$ is the 
transmission or spreading rate, $\beta>0$ is the removal rate and $N$ is the fixed population size, which implies that the model neglects the effects of spatial migration. Currently, there is no vaccination available for {\small COVID-19}, and the only way to reduce the transmission or infection rate  $\lambda$ -- which is often referred to as ``flattening the curve''-- is by implementing strong social distancing and hygiene measures.

The system is reduced to a first-order differential equation, which does not possess an explicit 
solution, but can be solved numerically. 
The SIR model can then be parametrized using actual infection data to solve $I(t)$, in order to investigate the evolution of the disease.
In the D model, we make the drastic assumption of \emph{no recovery} in order to obtain an analytical formula to describe -- instead of infestations -- 
the death evolution by {\small COVID-19}.  
This can be useful as a fast method to foresee the global behavior as a first approach, before applying more sophisticated methods. 
We shall see that the resulting D model describes well enough the data of the current pandemics in different countries.

\begin{figure*}[!ht]
\begin{center}
\includegraphics[width=7.cm,height=5.5cm,angle=-0]{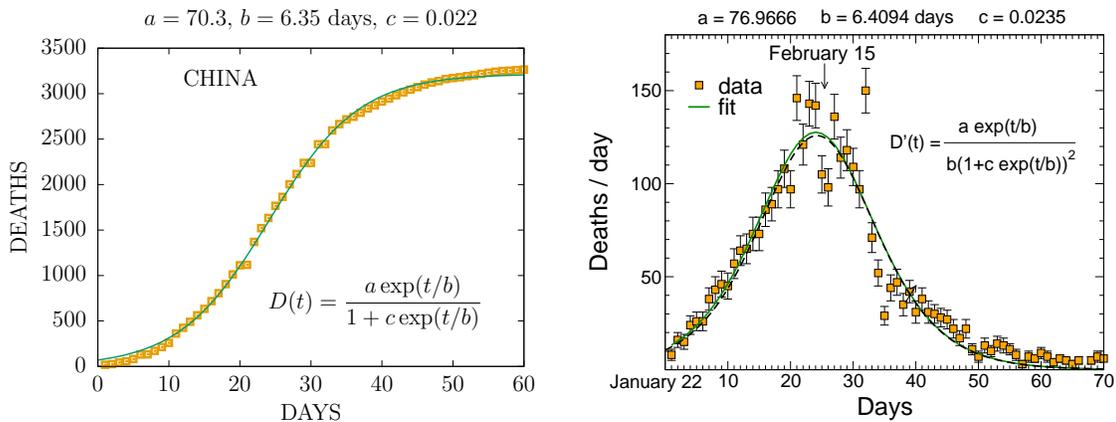}
\hspace{5mm}
\includegraphics[width=7.cm,height=5.5cm,angle=-0]{Fig1b.eps}
\caption{Fits to total (left panel) and daily (right panel) deaths by {\small COVID-19} in China using the $D(t)$ 
and D$^{\prime}$(t), respectively. 
The dashed curve shows a fit to the daily deaths using the parameters determined to fit the total deaths (top of left panel), 
which provides similar results to an independent fit (parameters on the top right), given the statistical fluctuations in the daily rates. 
Data are taken from \cite{worldometer}.\label{fig:chinaa}}
\end{center}
\end{figure*}

\section{The Death or D Model}

The main assumption of the D model is the absence of recovery from coronavirus, i.e. $R(t)=0$, at least during the pandemic time interval. 
This assumption may be reasonable if the spreading time of the pandemic is much faster than the recovery time, i.e.  $\lambda \gg \beta$. 
The SIR equations are then reduced to the single equation of the well-known SI model, 
\begin{equation}
  \frac{dI}{dt}  =   \lambda (N-I(t)) I(t),  \\
  \label{eq:SI}
\end{equation}
which represents the simplest mathematical form of all disease models, where the infection rate is proportional to both 
the infected, $I$, and susceptible individuals $N-I$.  Equation~\ref{eq:SI} is trivially solved by multiplying by $dt$ 
and dividing by $(N-I)I$,
\begin{equation}
\frac{dI}{(N-I)I} = \lambda dt,
\end{equation}
or
\begin{equation}
  \left(
  \frac{1}{N-I}
+  \frac{1}{I}
\right)dI
= \lambda N dt.
\end{equation}  
Integrating over an initial $t=0$ and final $t$ we obtain
\begin{equation}
  \ell n\frac{I(t)}{N-I(t)} - 
  \ell n\frac{I_0}{N-I_0} = \lambda N (t-t_0), 
\end{equation}
where $I_0=I(t_0)$. Taking the exponential on both sides
\begin{equation}
  \frac{I(t)}{N-I(t)} =   \frac{I_0}{N-I_0} {\rm e}^{\lambda N (t-t_0)}.
\end{equation}
Finally, solving this algebraic equation we obtain the solution $I(t)$
\begin{equation}
  I(t) = \frac{ N I_0  {\rm e}^{\lambda N (t-t_0)}  }
  {  N-I_0 +  I_0  {\rm e}^{\lambda N (t-t_0)}  }, 
\end{equation}
which can be written in the form
\begin{equation}
  I(t) =  \frac{ I_0 \,{\rm e}^{ (t-t_0)/b}  }
  {  1-C +  C \, {\rm e}^{(t-t_0)/b}  },
  \label{eq:It}
\end{equation}
where we have defined the constants
\begin{equation}  \label{param}
  b= \frac{1}{\lambda N}, \kern 1cm C = \frac{I_0}{N}.
  \end{equation}
The parameter $b$ is the characteristic evolution time of the initial
exponential increase of the pandemic. The constant $C$ is the initial
infestation rate with respect to the total population $N$. 
Assuming  $C \ll 1$, 
Eq.~\ref{eq:It} yields 
\begin{equation}
  I(t) =  \frac{ I_0 \,{\rm e}^{ (t-t_0)/b}  }
  {  1+  C \, {\rm e}^{(t-t_0)/b}  }.
  \label{eq:It2}
\end{equation}

In order to predict the number of deaths in the D model we assume that the
number of deaths at some time $t$ is proportional to the infestation
at some former time $\tau$, that is,
\begin{equation}
  D(t) = \mu I(t-\tau),
  \label{mu}
\end{equation}
where $\mu$ is the  death rate, and $\tau$ is the death time. 
With this assumption we can finally write the D-model equation as
\begin{equation} \label{D-model}
  D(t) =  \frac{ a {\rm e}^{ (t-t_0)/b}  }
  {  1 +  c \, {\rm e}^{(t-t_0)/b}  },
\end{equation}
where $a= \mu I_0 \, {\rm e}^{-\tau/b} $,  
 $c = C \, {\rm e}^{-\tau/b} $, and $a/c$ yields the total number of deaths predicted by the model.   
This is the final equation for the D-model, 
which presents a similar shape  to the well-known Woods-Saxon potential for the nucleons inside the atomic nucleus 
or the bacterial growth curve.  
The rest of the parameters, $\mu$, $\tau$, $I_0$ and $N$ are embedded in the parameters $a, b, c$, which represent space-time averages 
and can be fitted to the timely available data.

In Fig.~\ref{fig:chinaa}, we present the fit of the D-model to the {\small COVID-19} death data for China, where its evolution has apparently been controlled and the D function has reached the plateau zone, with few increments over time, or fluctuations that are beyond the model assumptions. This plot shows the duration of the pandemic -- about two months to reach the top end of the curve -- and the agreement, despite the crude assumptions, 
between data and the evolution trend described by the D-model. This agreement encourages the application of the D model 
to other countries in order to investigate the different trends.

\begin{figure}[!ht]
\label{parametros}
\begin{center}
\includegraphics[width=8.3cm,height=12.5cm,angle=-0]{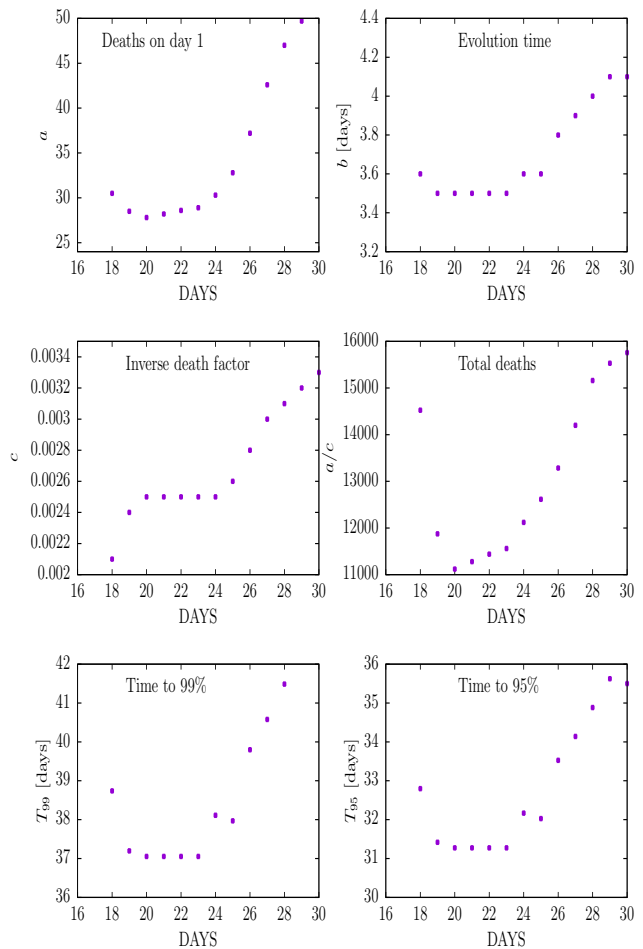}
\caption{Evolution of a, b  and c parameters and various predictions of the D-model 
as a function of time (days). 
}
\end{center}
\end{figure}

\subsection{Evolution of D-model parameters}

In order to get insight into the stability and uncertainty of our predictions, Fig. \ref{parametros} shows the evolution of 
$a$, $b$, and $c$ and other model predictions from fits to the daily data in Spain. The meaning of   
these quantities is explained below:

\begin{itemize}
  \item The parameter $a$ is the theoretical number of deaths at the day corresponding to $t=0$. 
    In general,  it differs from the experimental value and can be interpreted as the
    expected value of deaths that day. Note that  experimental
    data may be subject to unknown systematic errors and different counting methods.

  \item The parameter $b$, as mentioned above, is the characteristic
    evolution time. During the initial exponential behavior, it indicates the number of days for the number of deaths to double. 
    Moreover,  $1/b$ is proportional to the slope of the almost linear behavior in the mid
    region of the $D$ function. That behavior can be obtained
    by doing a Taylor expansion around $t_0=-b~\ell n~c$ and is given by 
    \begin{equation}
      D(t) \simeq \frac{1}{c}\left(1-\frac12\ell n ~c\right)+\frac{t}{2bc}.
      \end{equation}
      
  \item The parameter $c$ is called the 
    {\em inverse dead factor} because $D(t\rightarrow\infty)= a/c$ provides the asymptotic or expected total number of deaths. 


  \item The times $T_{95}$ and $T_{99}$ correspond to $D= 0.95
    D(\infty)$ and $D= 0.99 D(\infty)$, respectively. These times are
    obtained by solving the equation $D(t)= \gamma a/c$, where
    $\gamma=0.95$ or 0.99. The solution of that equation is
    \begin{equation}
      t= b~\ell n\left( \frac{1}{c}\frac{\gamma}{1-\gamma}\right).
    \end{equation}    
\end{itemize}

Figure~\ref{parametros} shows the stable trend of the parameters 
between days 19 to 24 (corresponding to March 27--30), right before 
reaching the peak  of deaths cases, which occurred in Spain around April 1. 
Such stability validates the D-model predictions during this time. 
However, a rapid change of the parameters is observed, especially for $a$, 
once the peak is reached, drastically changing the prediction of the number 
of deaths given by $a/c$. This sudden change results in the slowing down of deaths per day and 
longer time predictions $T_{95}$ and $T_{99}$. 

The parameters of the D model 
correspond to average values over time of the interaction coefficients between individuals, i.e. they are sensitive 
 to an additional external effect on the pandemic 
evolution. These may include the  lockdown effect imposed in Spain in March 14 and other effects such as new sources of 
infection or a sudden increase of the total susceptible individuals due to social migration and large mass gatherings~\cite{hunter}. 
It is not possible to identify a specific cause because its effects are  blurred by the stochastic evolution of the pandemic, 
which is why any reliable forecast presents large errors.

\subsection{The $D^{\prime}$ model}

One can also determine deaths/day rates by applying the first derivative to Eq.~\ref{D-model}, 
\begin{equation} \label{Dprime-model}
  D^{\prime}(t) =  \frac{ a {\rm e}^{ (t-t_0)/b}  }{ b (1 +  c \, {\rm e}^{(t-t_0)/b})^2  },
\end{equation}
which allows for a determination of the pandemics peak and evolution after its turning point. 
The $D$ model describes well the cumulative deaths because the sum of
discrete data reduce the fluctuations, in the same way as the integral of a discontinuous function  is a continuous function.
However, the daily data required for $D^{\prime}$ have large fluctuations -- both statistical and systematic -- which 
normally gives a slightly different set of parameters when compared with the D  model.

\begin{figure}[!ht]
\begin{center}
\includegraphics[width=8.3cm,height=4.5cm,angle=-0]{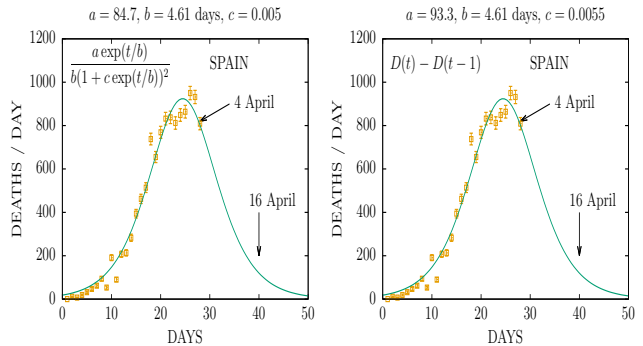}
\caption{Predictions of the D model for $D^{\prime}(t)$ in Spain according with
  the data collected up to April 5.\label{diferencia}
}
\end{center}
\end{figure}

\begin{figure}[!ht]
\begin{center}
\includegraphics[width=4.1cm,height=4.5cm,angle=-0]{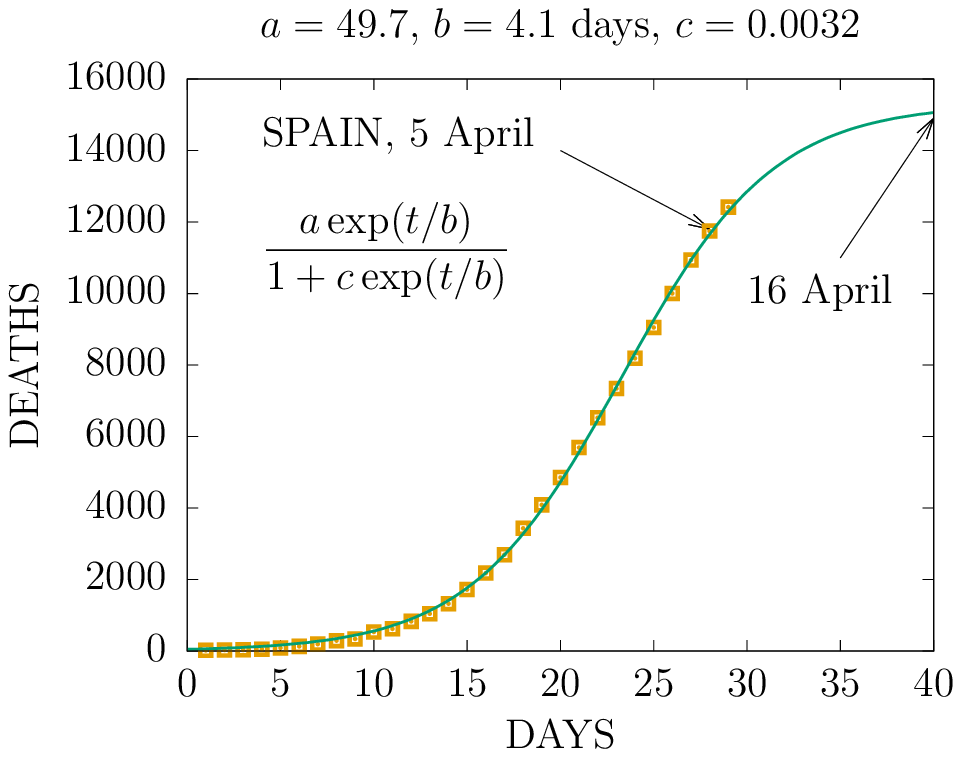}
\includegraphics[width=4.1cm,height=4.5cm,angle=-0]{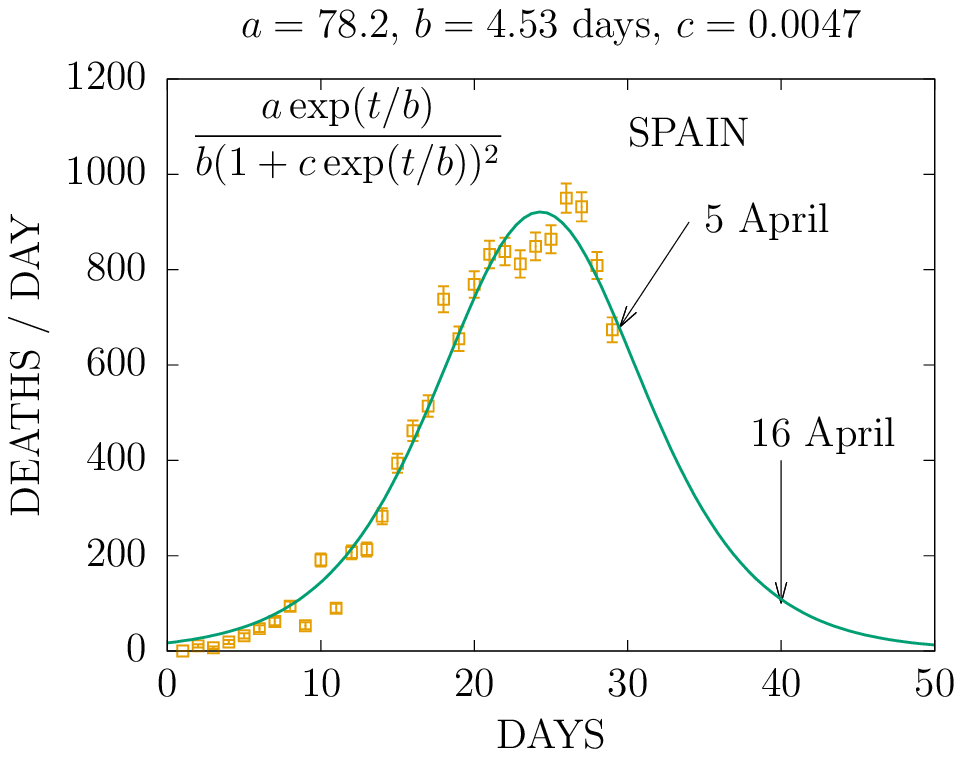}
\caption{Comparison of parameters fitted to $D(t)$ and $D^{\prime}(t)$ in Spain according with
  the data on April 5. \label{meseta}}
\end{center}
\end{figure}

Using the $D$ model fitted to cumulative deaths allows to compute deaths/day as
\begin{equation}
  D(t)-D(t-\Delta t) \simeq D^{\prime}(t) \Delta t,
  \label{Delta}
\end{equation}
where $\Delta t=1$ day. Figure \ref{diferencia} shows that Eqs. \ref{Dprime-model} and \ref{Delta} yield similar parameters, 
as the time increment is small enough compared with the time evolution of the $D(t)$ function.  
Hence, the first derivative $D^{\prime}(t)$ can be used to describe deaths per day. In addition, Fig.~\ref{meseta} shows that the 
parameters may be different for both $D$ and $D^{\prime}$ functions using 
cumulative and daily deaths, respectively, as shown for Spain on April 5.  
It is also important to note that $b$ is directly proportional to the full width at half maximum ($FWHM$) of the $D^{\prime}(t)$ distribution,
\begin{equation}
FWHM= 2b ~\ell n (3+2\sqrt{2}) \approx 3.5~b. 
\end{equation}
As shown below, the $b$ parameter presents typical values between 4 and 10 for most countries undergoing the initial exponential phase, 
which yields a minimum and maximum time of 14 and 35 days, respectively, between the two extreme values of the $FWHM$.

\subsection{$D_n$ model with two or more channels of infection}

\begin{figure}[!ht]
\begin{center}
\includegraphics[width=7cm,height=5.5cm,angle=-0]{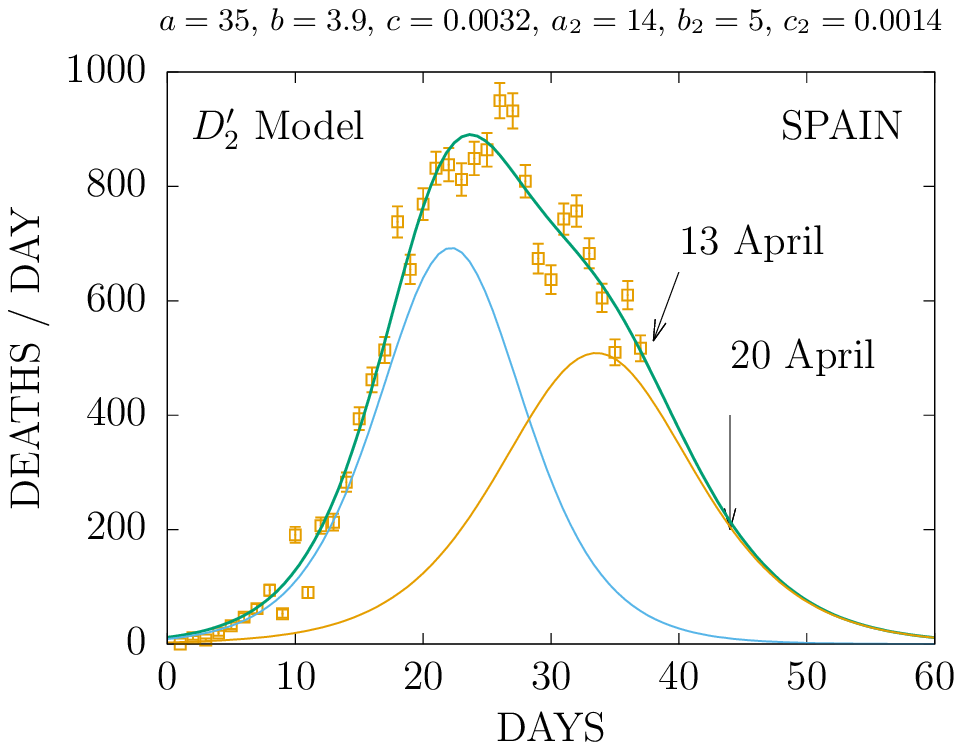}
\caption{Predictions of the $D^{\prime}_2$ model in Spain using a sum of two
  D$^{\prime}$-functions for data collected up to April 13. \label{doble}}
\end{center}
\end{figure}

Some models \cite{Deh20} include changes in the transmission rate due to various interventions implemented to contain the outbreak. 
The simple D model does not allow to do this explicitly, but changes in the spread can be taken into account 
by considering the total D or $D_n$ function as the sum of two or more independent D-functions with different parameters, 
which may reveal the existence of several independent sources, or virus channels.
An example is shown in Fig. \ref{doble}, where the two-channel function
\begin{equation}
 D^{\prime}_{\rm 2} =  D^{\prime}(a,b,c)+D^{\prime}(a_2,b_2,c_2), 
\end{equation}  
has been fitted with six parameters to the Spanish data up to April 13. The fit reveals a second, smaller 
death peak, which substantially increase the number of deaths per day and the duration of the pandemic. 
This is equivalent to add a second, independent, source of infection
several weeks after the initial pandemic. The second peak may as well represent a second pandemic 
phase driving the effects of quarantine during the descendant part of the curve.


\begin{figure}[!ht] 
\begin{center}
\includegraphics[width=7cm,height=5.5cm,angle=-0]{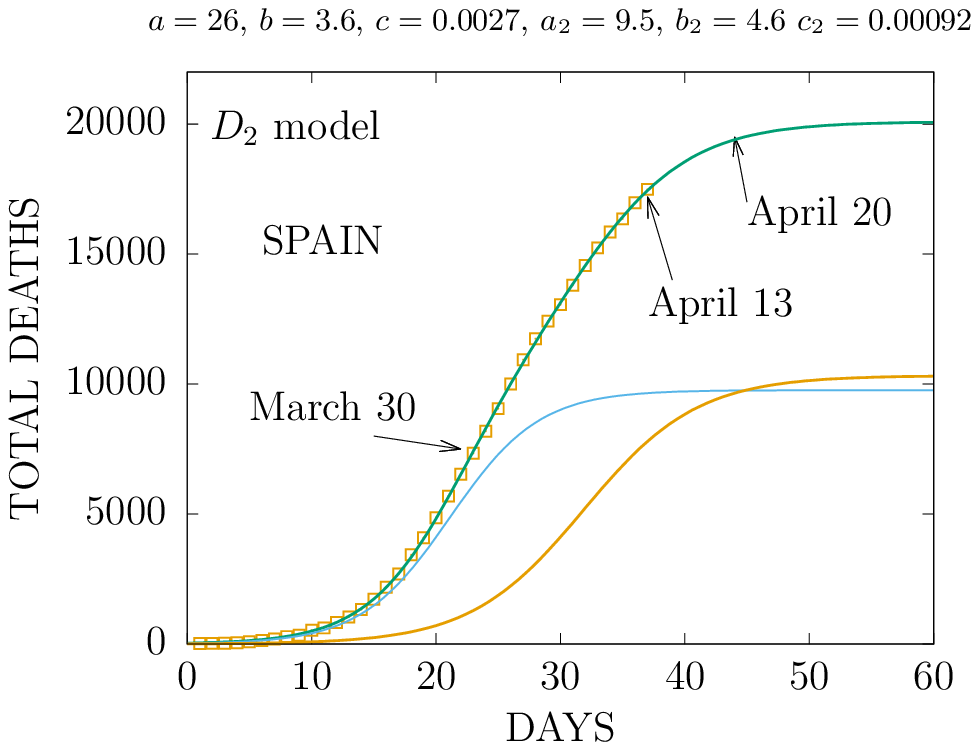}
\caption{Predictions of the $D_2$ model  in Spain using a sum of two
  D-functions for data collected up to April 13.\label{ddmeseta}
 }
\end{center}
\end{figure}

Additionally, the cumulative  D-function can also be computed with a two-channel function, 
\begin{equation}
 D_{\rm 2} =  D(a,b,c)+D(a_2,b_2,c_2), 
\end{equation}  
which provides, as shown in Fig. \ref{ddmeseta}, a more accurate prediction for the total number of deaths 
and clearly illustrates the separate effect of both source peaks. It is interesting to note that for large $t$, 
$a\approx a_2$, $c\approx c_2$ and $b_2\approx 2b$. In such a case, the total number of deaths expected during the pandemic 
is given by $D_2(\infty)=2a/c$.

\subsection{Estimation of the infected function $I(t)$}

The D-model can also be used to estimate $I(t)$ using the initial values of $I_0=I(0)$ and the total number of susceptible people
$N=S(0)$. The initial value of $N$ is unknown, and not necessarily equal to the population of the whole country since the pandemic started in localized areas.
Here, we shall assume 
$N=10^6$, although plausible values of $N$ can be tens of millions.  
Note that the no-recovery assumption of the D model is unrealistic, and 
this calculation only provides an estimation of the number of individuals that were 
infected at some time, independently of whether they recovered or not.

\begin{figure}[!ht] 
\begin{center}
\includegraphics[width=7cm,height=5.5cm,angle=-0]{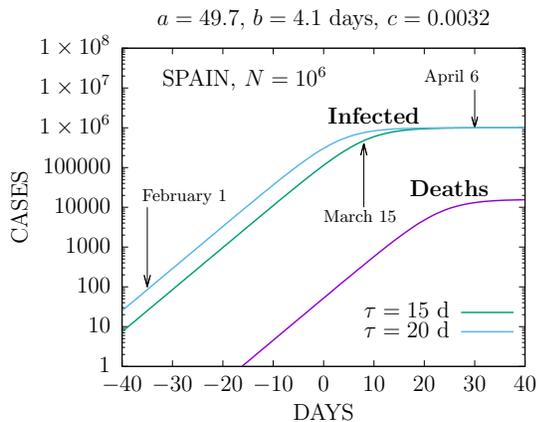}
\caption{Predictions of the D  model for the infected function $I(t)$ in Spain according to data collected up to April 6. 
\label{infected}
}
\end{center}
\end{figure}

From the definition of $D(t)$ in Eq.~\ref{mu}, the following relations between the several parameters of the model were 
extracted
\begin{eqnarray}
a & = & \mu I_0 {\rm e}^{-\tau/b}, \\
c & = & \frac{ I_0}{N} {\rm e}^{-\tau/b}, \\
b &=& \frac{1}{\lambda N}.
  \end{eqnarray}
Solving the first two equations for $\mu$ and
$I_0$ we obtain
\begin{eqnarray}
  I_0 &=& N c \, {\rm e}^{\tau/b}, \\
  \mu &=& \frac{a}{Nc}.
\end{eqnarray}
Hence, $\mu$ can be computed by knowing $N$. However, to obtain $I_0$ one needs to
know the death time $\tau$. This has been estimated to be about 15 to
20 days for {\small COVID-19} cases, which can be used to compute two estimates of
$I(t)$. These are given in Fig.~\ref{infected} for the case of Spain.

\begin{figure}[!ht] 
\begin{center}
\includegraphics[width=6.5cm,height=5cm,angle=-0]{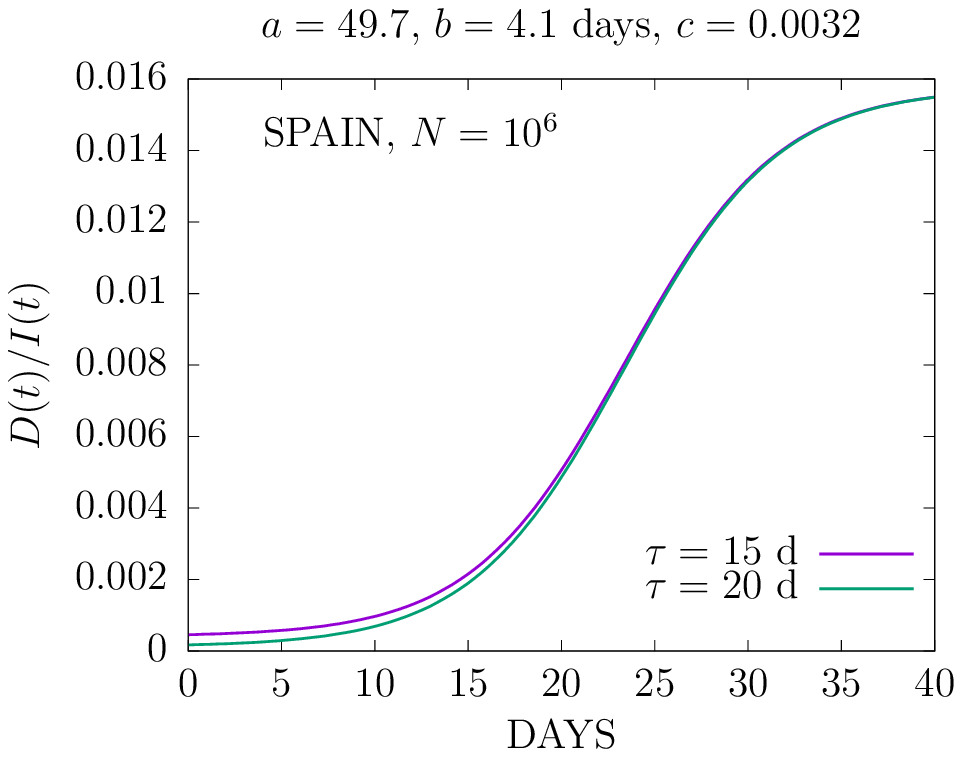}\\
\vspace{0.3cm}
\includegraphics[width=6.5cm,height=5cm,angle=-0]{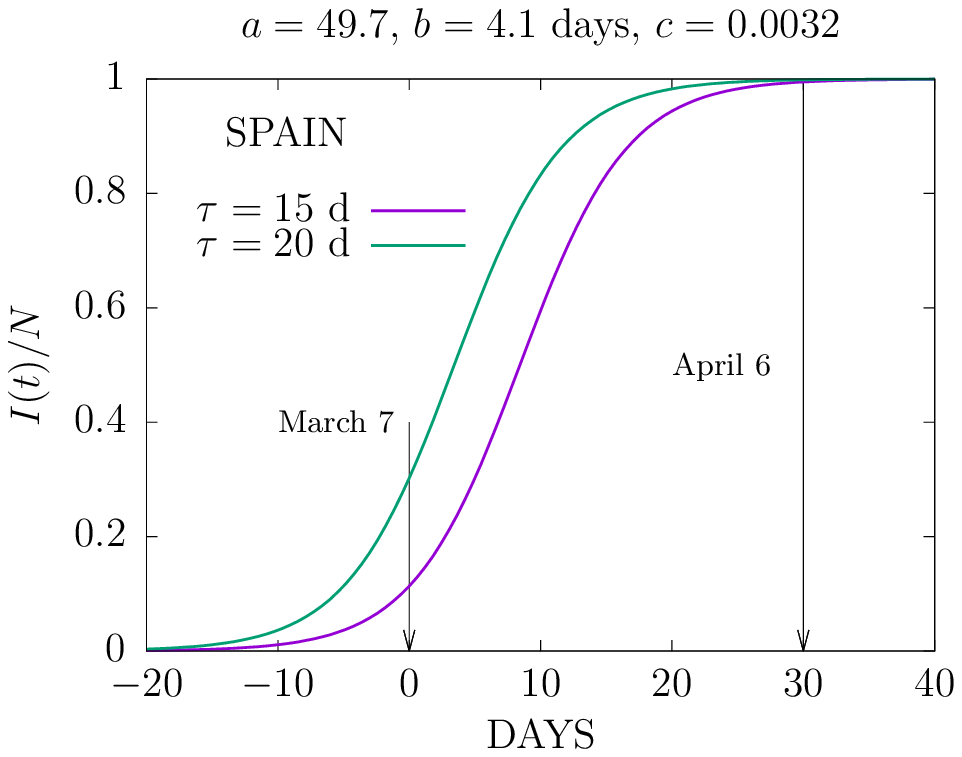}
\caption{(Top panel) Results for  $D(t)/I(t)$ (deaths over infected) and (bottom panel)  $I(t)/N$ (infected over susceptible) 
according to the D-model. Results are for data collected in Spain up to April 6, assuming $N=10^6$. \label{di}}
\end{center}
\end{figure}

Since there is no recovery in the D model, the total number of infected people 
is $I\sim N$ for large $t$, i.e. $N=10^6$ in our case.  In Fig.~\ref{infected}, 
we have labeled the beginning of the lockdown in Spain (March 15). For $\tau=15$ days, most of the susceptible individuals were 
already infected on that date, and even more for $\tau=20$ days, as the pandemic had started almost two months earlier. 
Most of the individuals got infected, even if a great part of them -- approximately 99\% -- had no symptoms of illness or disease.

Moreover, the top panel of Fig.~\ref{di} shows the  ratio $D(t)/I(t)$ (deaths over infected), as given by 
Eqs.~\ref{eq:It2} and \ref{D-model},
\begin{equation}
  \frac{D(t)}{I(t)}=
  \frac{a}{ Nc\, {\rm e}^{\tau/b} } 
    \frac{1+c\, {\rm e}^{ (t+\tau) /b } }{1+c\, {\rm e}^{\tau/b}},
\end{equation}
which also depends on $N$ and $\tau$. For $N=10^6$, the ratio $D/I$ increases similarly to the separate 
functions $D$ and $I$ between the initial and final values,
\begin{eqnarray}
 \frac{D(0)}{I(0)} &=&  \frac{a}{Nc\, {\rm e}^{\tau/b}},\\
 \frac{D(\infty)}{I(\infty)} &=&  \frac{a}{Nc}.
\end{eqnarray}

\begin{figure*}[!ht] 
\begin{center}
\includegraphics[width=5.5cm,height=5.5cm,angle=-0]{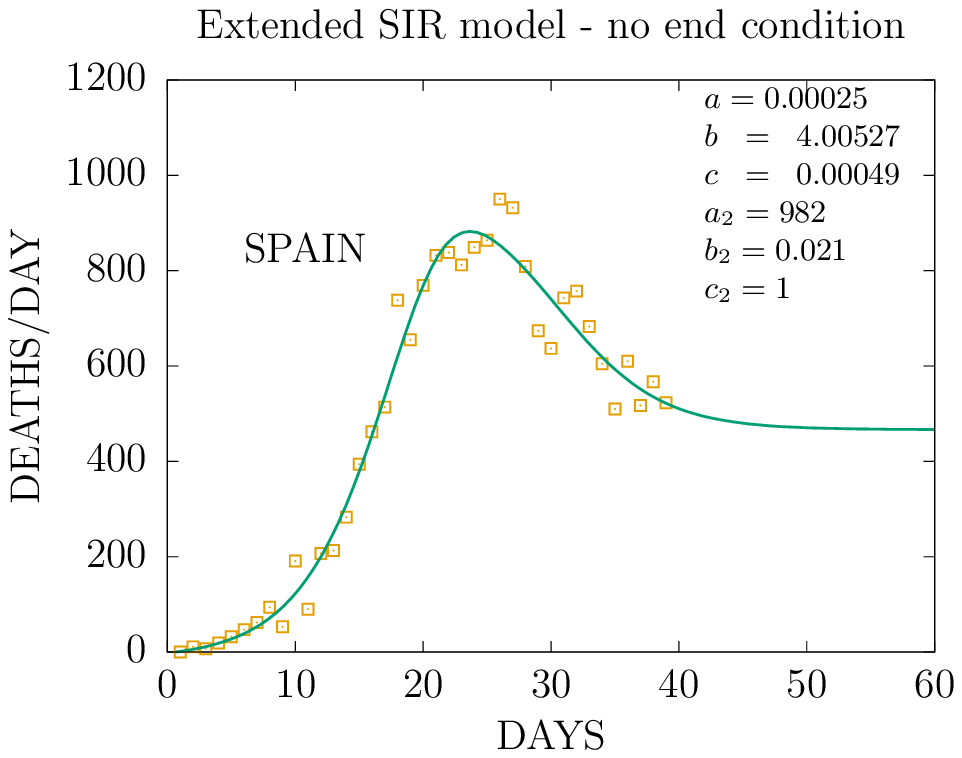}
\includegraphics[width=5.5cm,height=5.5cm,angle=-0]{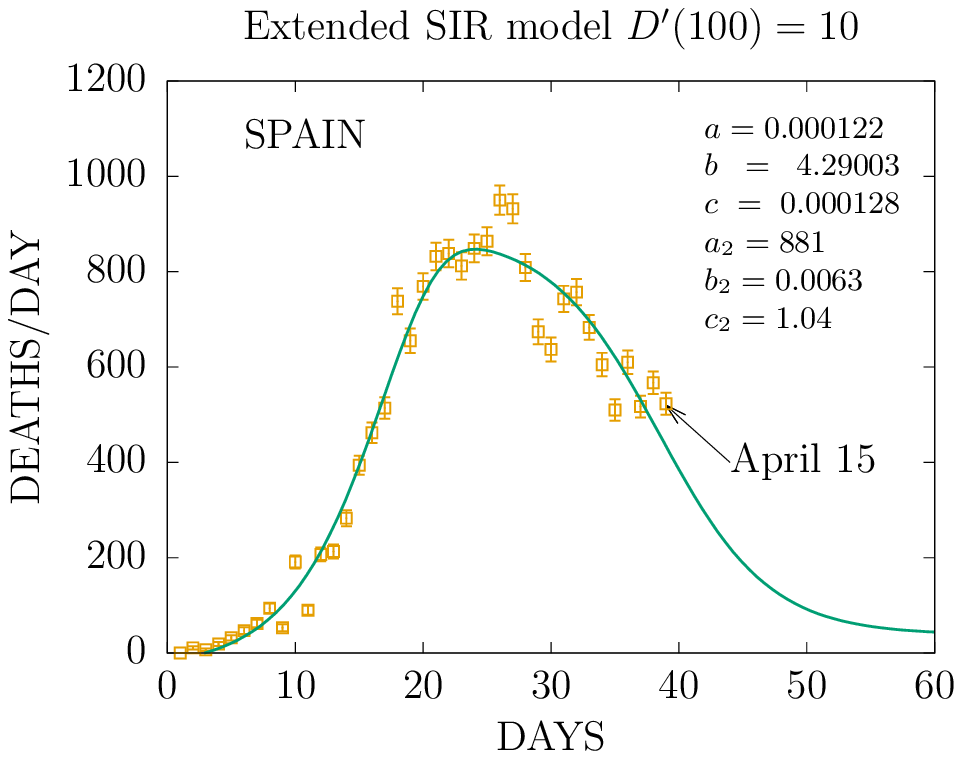}
\includegraphics[width=5.5cm,height=5.5cm,angle=-0]{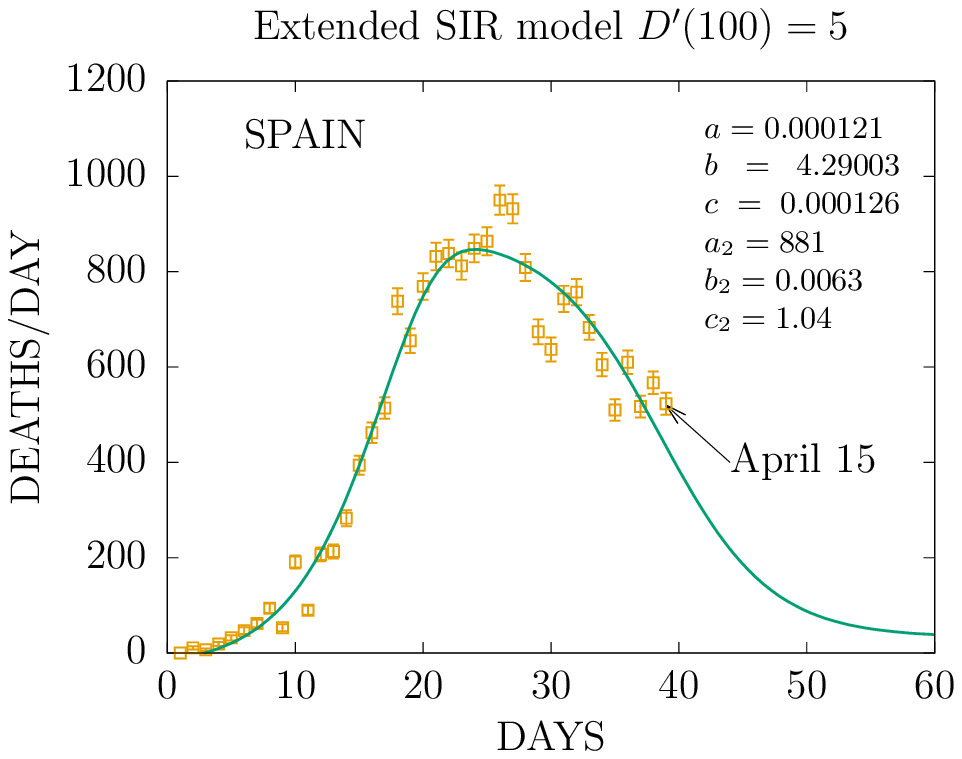}
\caption{Fit of the ESIR model to daily deaths in Spain up to April 15 using no boundary condition 
for the final number of deaths (left panel), and with boundary conditions of $D^{\prime}(100)=10$  (middle panel) and $D^{\prime}(100)=5$ (right panel) deaths/day.
\label{sirfig1}
}
\end{center}
\end{figure*}

These results depend on the total susceptible population $N$.
However,  the ratio of infected with respect to susceptibles, $I/N$, is independent on $N$. This
function depends only on $\tau$ and is shown in the bottom panel of Fig. \ref{di} for $\tau=15$ and 20 days, 
which reveals the rapid spread of the pandemic. Accordingly, 
between 10\% and 30\% of the susceptibles were infected in March 7, and one month later (April 6), 
when the fit was made, all susceptibles had been infected. 
This does not means that the full population of the country got infected, since the number $N$ is unknown 
and, for instance, excludes individuals in isolated regions, and it may additionally change because of spatial migration, not considered in the model.


\section{The extended SIR model}

D-model predictions can be compared with more realistic results given by the complete SIR model \cite{KM,bartlett1957}, 
which is  characterized by Eqs.~\ref{eq:10}, \ref{eq:11}, \ref{eq:12} and \ref{eq:13} with  
initial conditions $R(0)=0$, $I(0)=I_0$, $S(0)=N-I_0$.  The SIR system of dynamical equations 
can be reduced to a non-linear differential equation. 
First, dividing Eq.~\ref{eq:10} by Eq.~\ref{eq:12} one obtains, 
\begin{equation}
  \frac{dS}{dR}  =  - \frac{\lambda}{\beta} S,
\end{equation}
which yields the following exponential relation between the 
susceptible and the removed functions, 
\begin{equation}
  S= S_0 {\rm e}^{-\lambda R / \beta}.
\end{equation}
Moreover, Eq.~\ref{eq:13} provides a relation between the infected and the removed functions,
\begin{equation}
  I = N - S - R = N- S_0 {\rm e}^{-\lambda R / \beta}-R,
\end{equation}
which yields, by inserting into Eq.~\ref{eq:12}, the final SIR differential equation
\begin{equation}
\frac{dR}{dt} 
  = \beta \left( N- S_0 {\rm e}^{-\lambda R / \beta}-R \right).
\end{equation}

In order to obtain  $R(t)$ we only need to solve this first-order differential equation with the initial condition
$R(0)=0$. Moreover, if we normalize the functions $S$, $I$ and $R$ to 1,
\begin{eqnarray}
  S&=& s N, \\
  I&=& i N, \\
  R&=& r N,
\end{eqnarray}
so that $s+i+r=1$, then $r(t)$ verifies
\begin{equation} \label{sirequation}
\frac{dr}{dt} 
  = \beta \left( 1- s_0 {\rm e}^{-\lambda N r / \beta}-r \right),
\end{equation}
which can be solved numerically, or by approximate methods in
some cases. In Ref.~\cite{KM}, a solution was found for small values of
the exponent $\lambda N r / \beta$. For the coronavirus pandemic, however, 
this number is expected to increase and be close to one at the pandemic end. 

At this point, we propose a modification of the standard SIR model.
Instead of solving  Eq.~\ref{sirequation} numerically and fitting the
parameters to data,  the solution can be parametrized as
\begin{equation}\label{rfunction}
  r(t) = \frac{a}{c + {\rm e}^{-t/b}},
\end{equation}
which presents the same functional form as the D-model and, conveniently, provides a faster way to fit the model parameters 
by avoiding the numerical problem of solving Eq. \ref{sirequation}. In fact, numerical solutions of the SIR model 
present a similar step function for $R(t)$.
Additionally, one can assume that $D(t)$ is proportional to $R(t)$, and  can also be written as
\begin{equation}
\frac{dD}{dt} =  
   a_2 \left( 1- c_2 {\rm e}^{-r / b_2}-r(t)\right),
\end{equation}
where $a_2$, $c_2 = s_0$ and $b_2 = \beta/(\lambda N)$ are unknown parameters to be fitted to deaths-per-day data, together with the three parameters of the $r(t)$-function: $a$, $b$, $c$.  

\begin{figure*}[!ht] 
\begin{center}
\includegraphics[width=8cm,height=5cm,angle=-0]{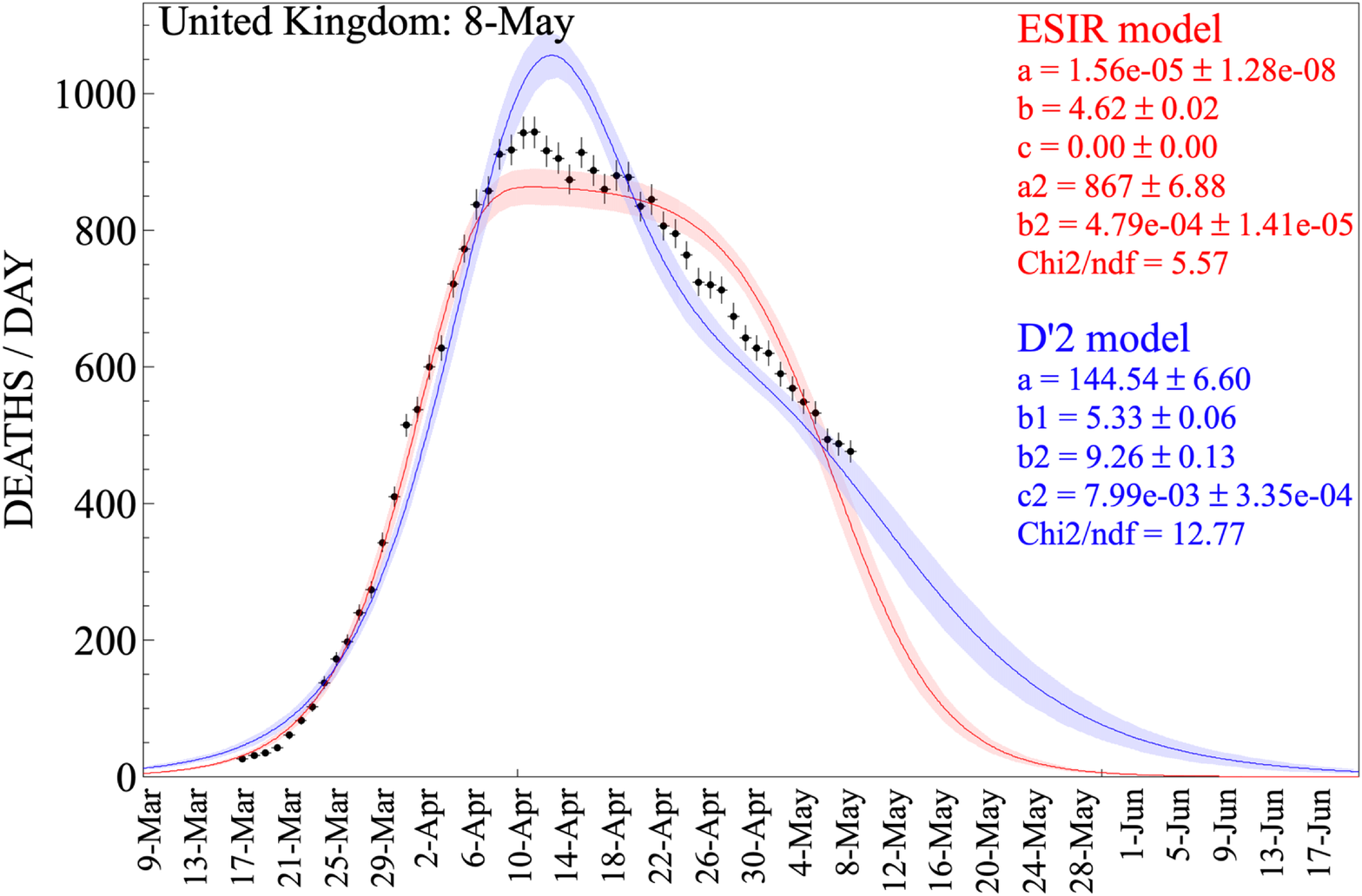}
\hspace{0.2cm} 
\includegraphics[width=8cm,height=5cm,angle=-0]{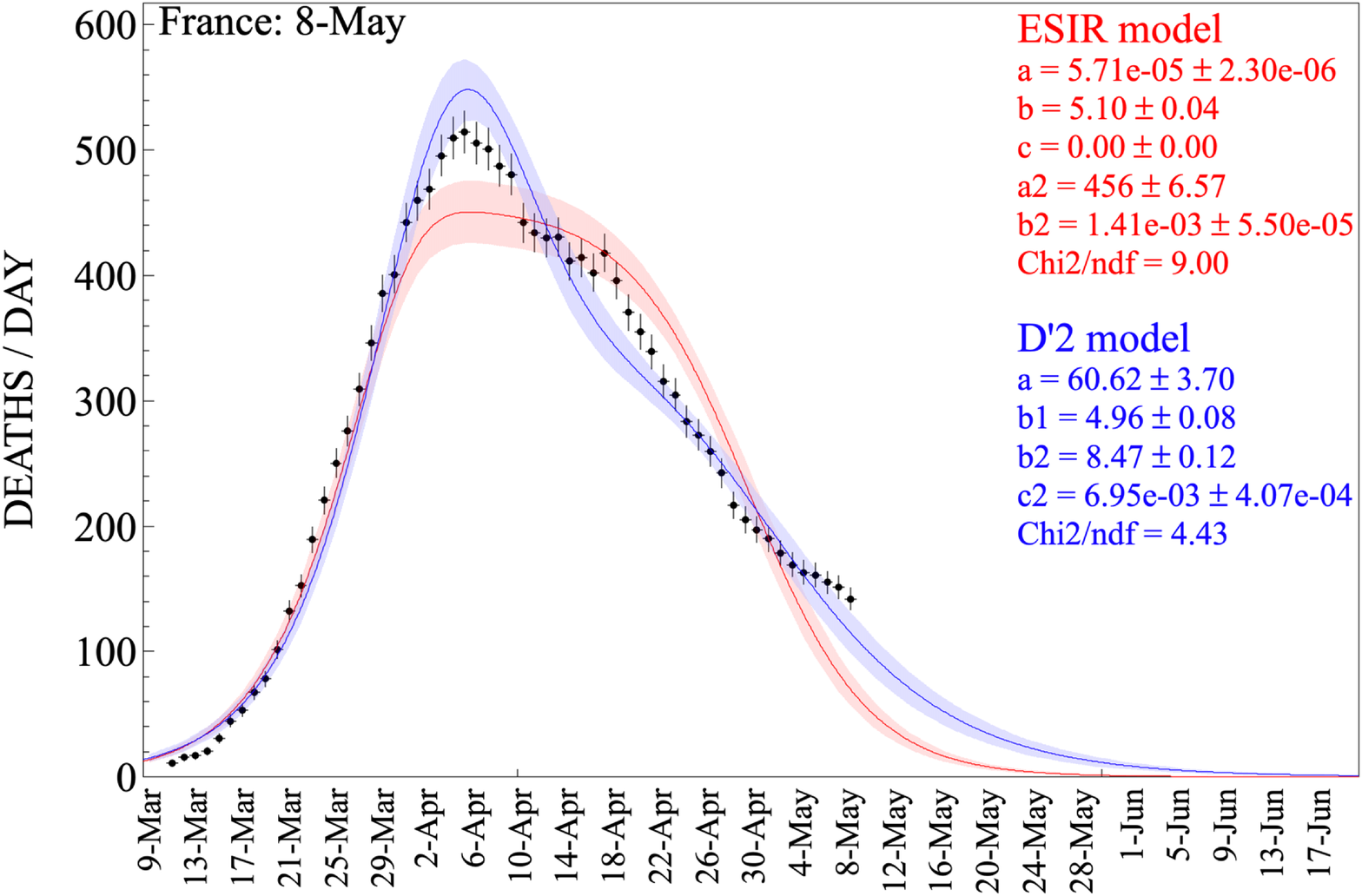}
\caption{Fit to the data (average of 7 consecutive days up to May 8) of the ESIR and $D^{\prime}_2$ models in the United Kingdom (left) 
and France (right).\label{esir12}
 }
\end{center}
\end{figure*}

Figure~\ref{sirfig1} shows fits of the ESIR model to daily deaths
in Spain during the coronavirus spread. The use of no boundary condition for the number of deaths 
(left panel) is  not an exact solution of the SIR differential equation. A way to solve this problem is to impose
the condition $D^{\prime}(\infty)=0$, as the number of deaths must stop at some time. 
Numerically, it is enough to choose a small value of $D^{\prime}(t)$  for an arbitrary large $t$.
The middle and right panels of Fig.~\ref{sirfig1} show different boundary conditions of 
$D^{\prime}(100)=10$ and $D^{\prime}(100)=5$, respectively, which yield  the same results 
and the expected behavior for a viral disease spreading and declining.

It is also consistently observed (e.g. see middle and right panels of Fig.~\ref{sirfig1}), that at large $t$, $r(t) \rightarrow \frac{a}{c} \approx 1$, which essentially means that most of the susceptible population $N$ recovers, as we previously inferred from the $D$ model. 
This, together with the fact that $c_2$ can always be adjusted to 1, leaves the ESIR model with essentially 
4 free parameters to fit to the daily death data; i.e. the same number of parameters than the original SIR model. 
As shown in Fig.~\ref{esir12},  ESIR fits reproduce well the long flattening behavior 
observed in UK, USA, Germany or Iran, whereas it fails to reproduce the more-pronounced 
double-peak structure typically observed in countries like France, Italy, Spain or Belgium.

As previously done with the D model, one can also expand the ESIR model to accommodate this apparent failure to 
take lockdown effects into account. Similarly, the ESIR2 model is proposed as, 

\begin{equation}
\mbox{ESIR2}(t) =   a_2 \left( 1- c_2 {\rm e}^{-r / b_2}-r(t)\right),
 \label{r2}
\end{equation}
with
\begin{eqnarray}
r(t)&=&\frac{a}{c+e^{-t/b}} + \frac{a^{\prime}}{c^{\prime}+e^{-t/b^{\prime}}} \nonumber  \\
&=& \frac{a}{2a+e^{-t/b}} + \frac{a}{2a+e^{-t/b^{\prime}}},
\label{r22}
\end{eqnarray}
where we have assumed that $a=a^{\prime}$ and $c=2a$ to accommodate that $r(\infty)\rightarrow 1$ 
and $c_2=1$. Hence, we are left with five free parameters.

\begin{figure*}[!ht]
\begin{center}
\includegraphics[width=4.5cm,height=8.3cm,angle=-90]{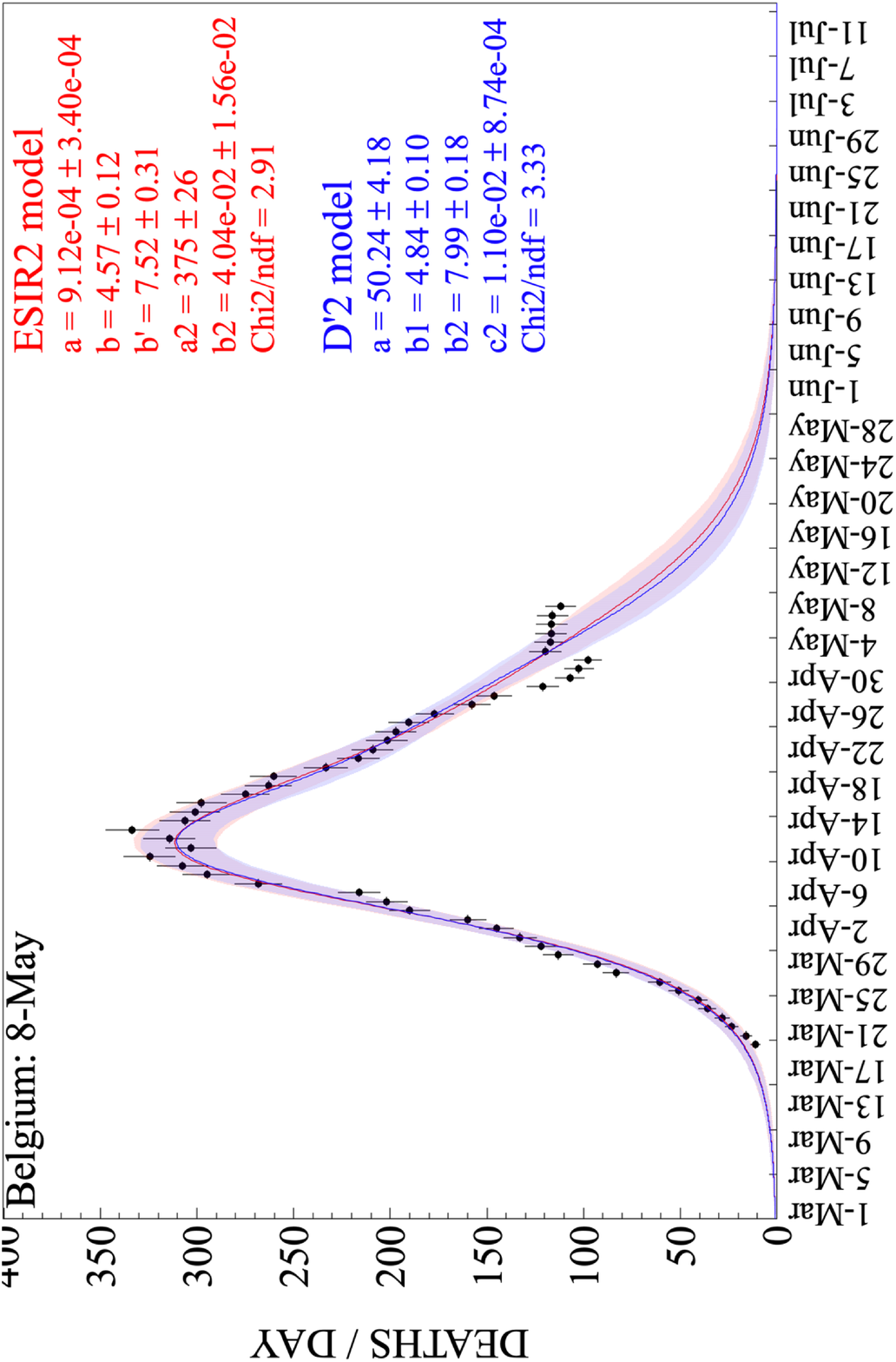}
\vspace{0.2cm} \hspace{0.3cm}
\includegraphics[width=4.5cm,height=8.3cm,angle=-90]{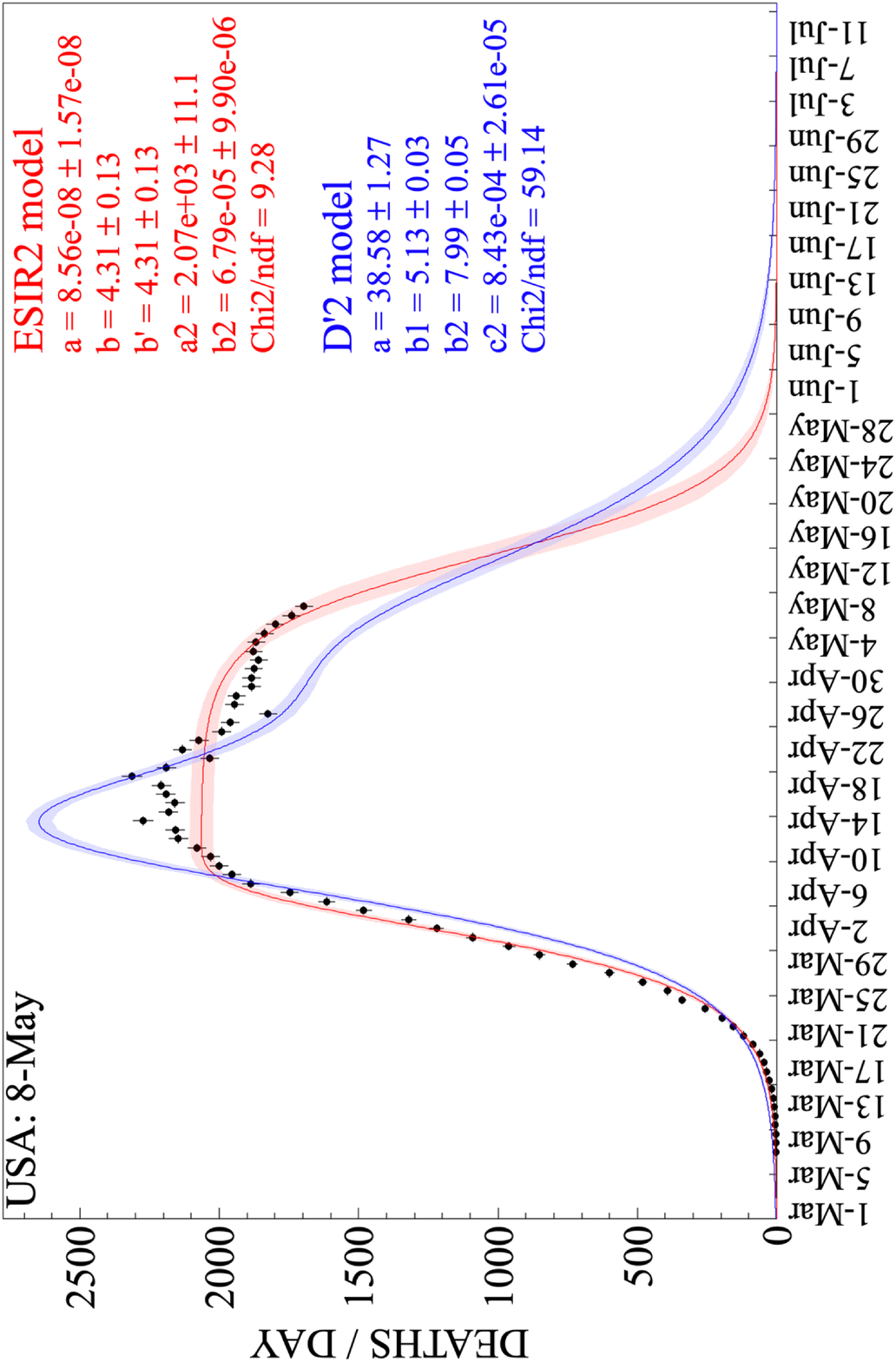}\\
\vspace{0.2cm} 
\includegraphics[width=4.5cm,height=8.3cm,angle=-90]{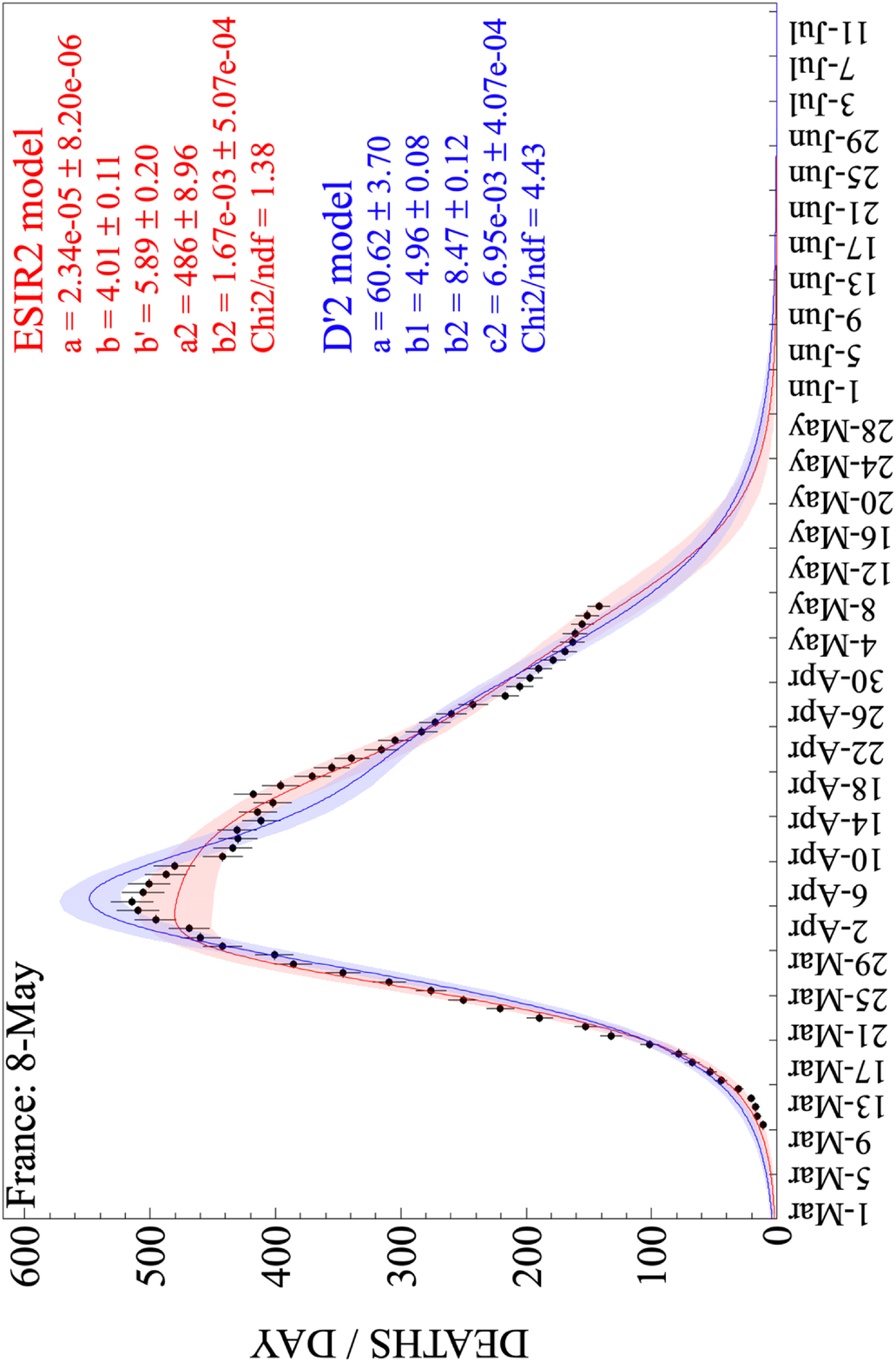}
\vspace{0.2cm} \hspace{0.3cm}
\includegraphics[width=4.5cm,height=8.3cm,angle=-90]{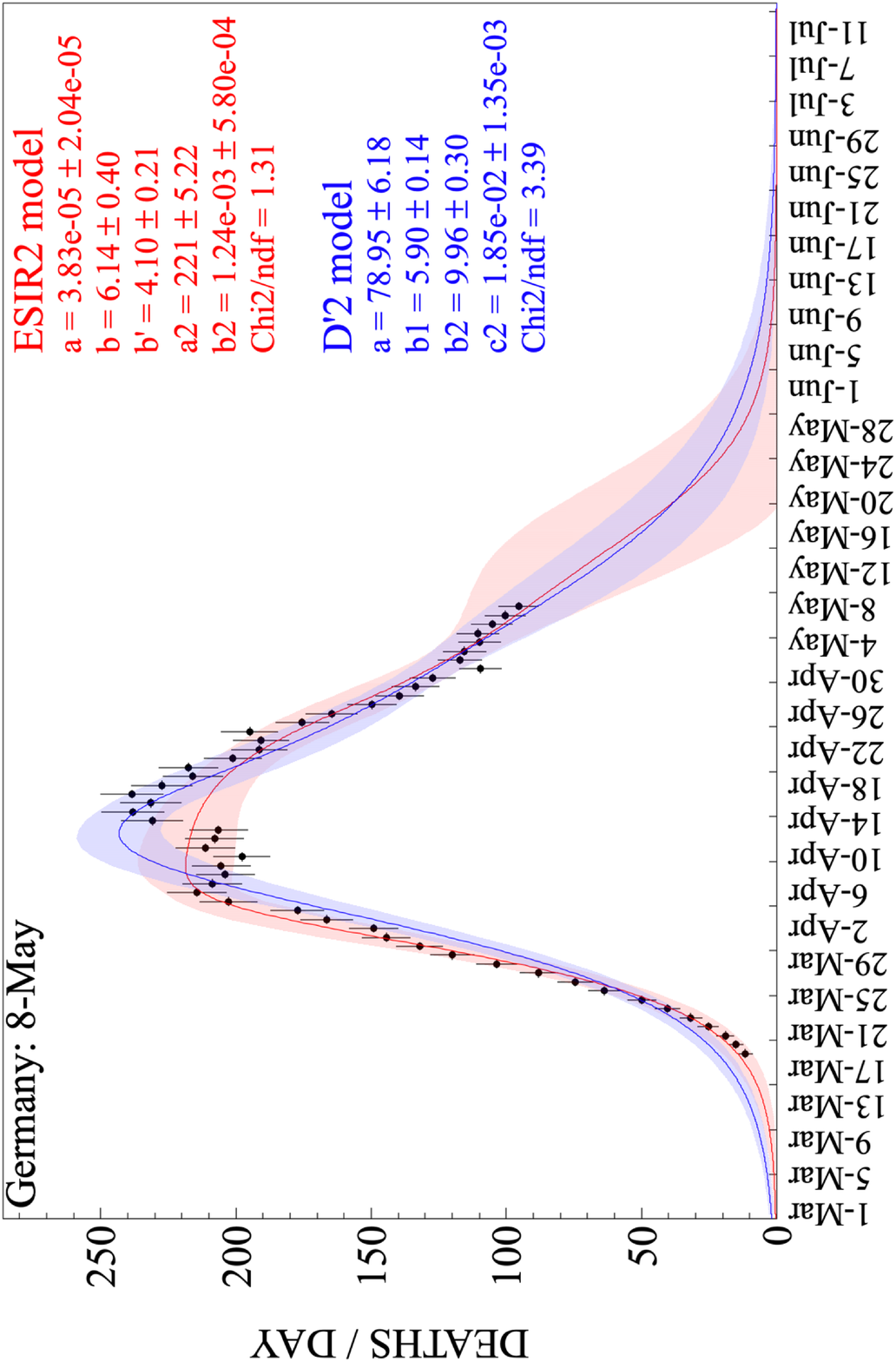}\\
\vspace{0.2cm} 
\includegraphics[width=8.3cm,height=4.5cm,angle=0]{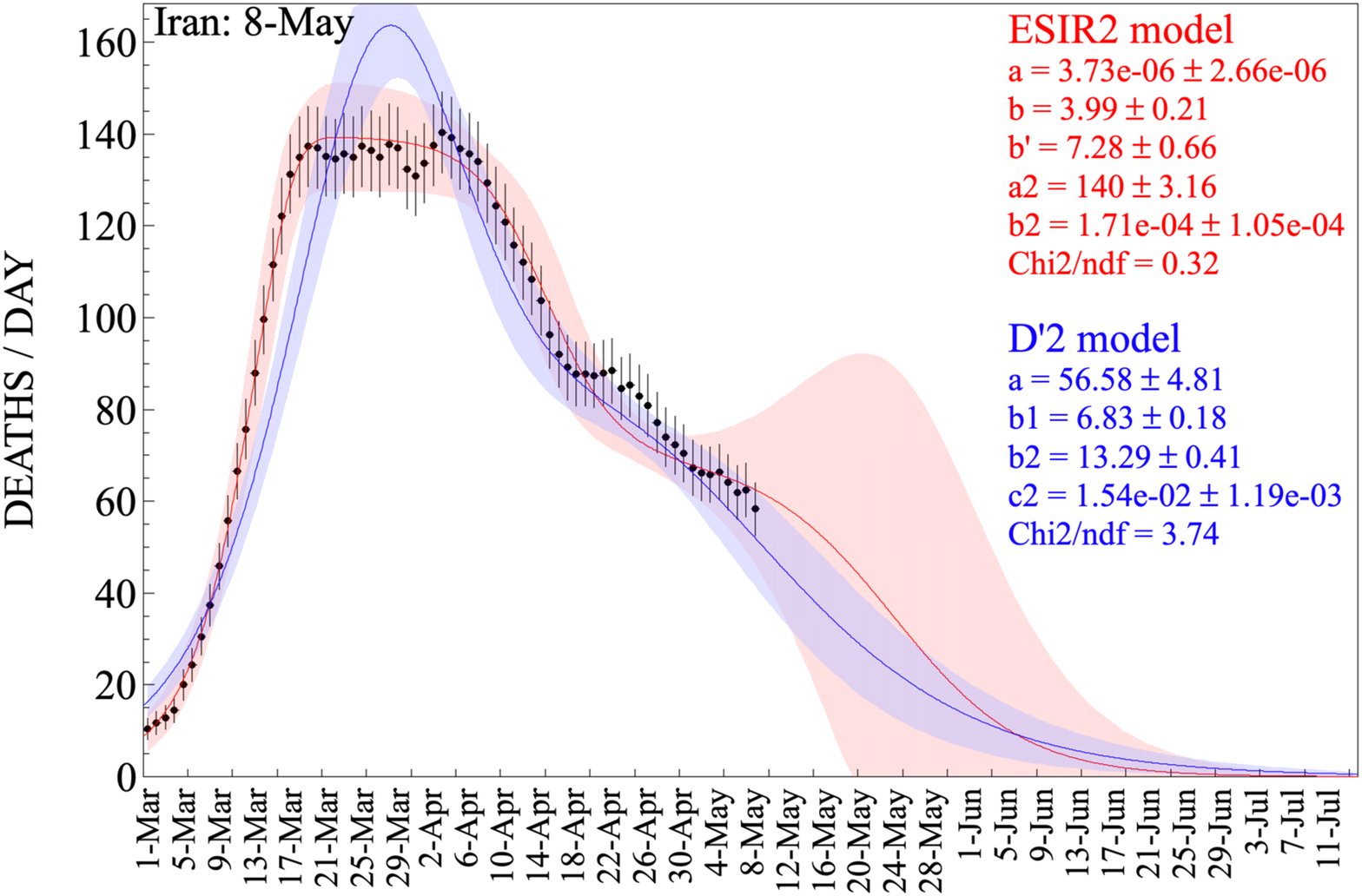}
\includegraphics[width=4.5cm,height=8.3cm,angle=-90,origin=br]{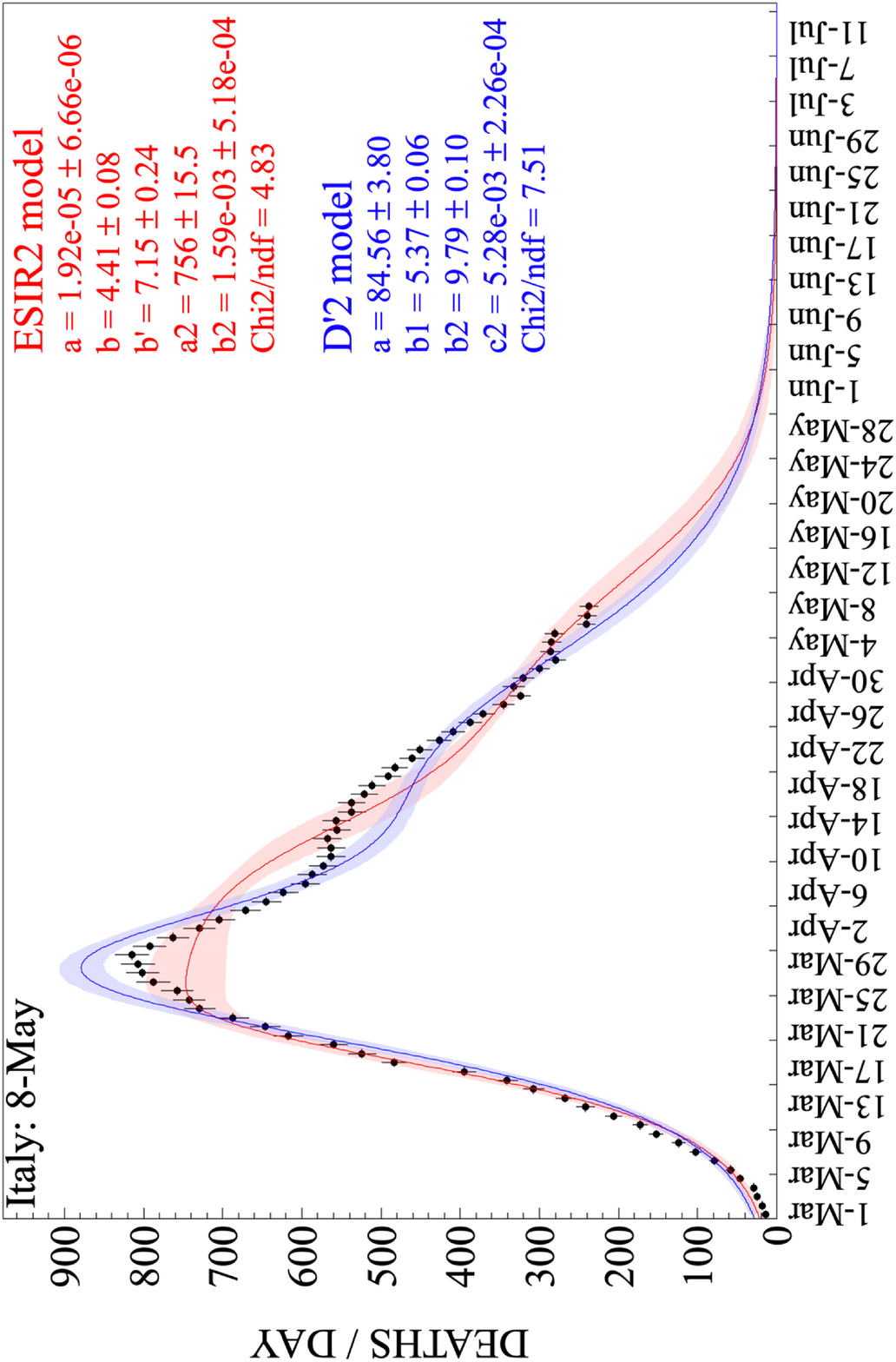}\\
\vspace{0.2cm} 
\includegraphics[width=4.5cm,height=8.3cm,angle=-90]{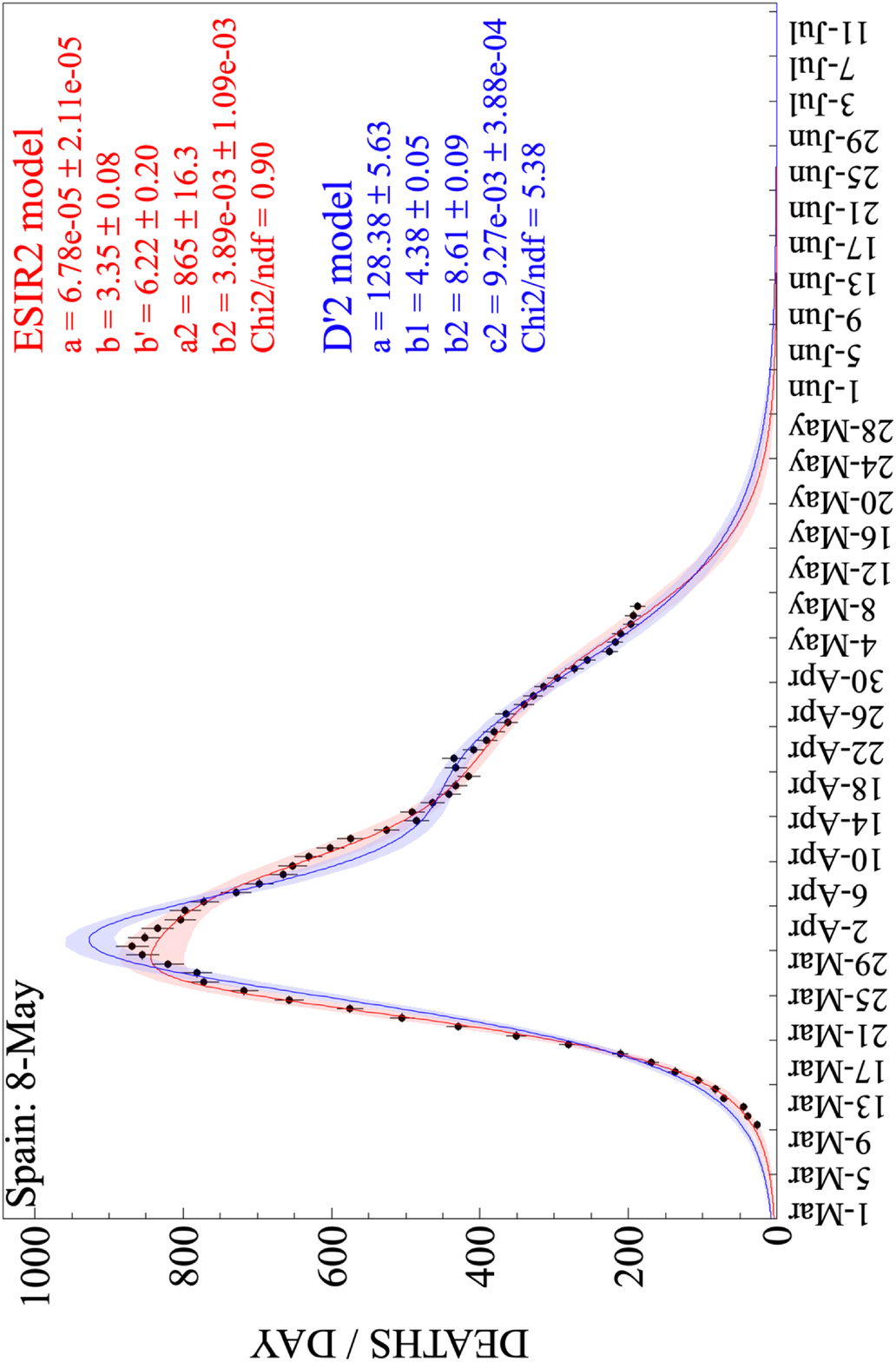}
\vspace{0.2cm} \hspace{0.3cm}
\includegraphics[width=4.5cm,height=8.3cm,angle=-90]{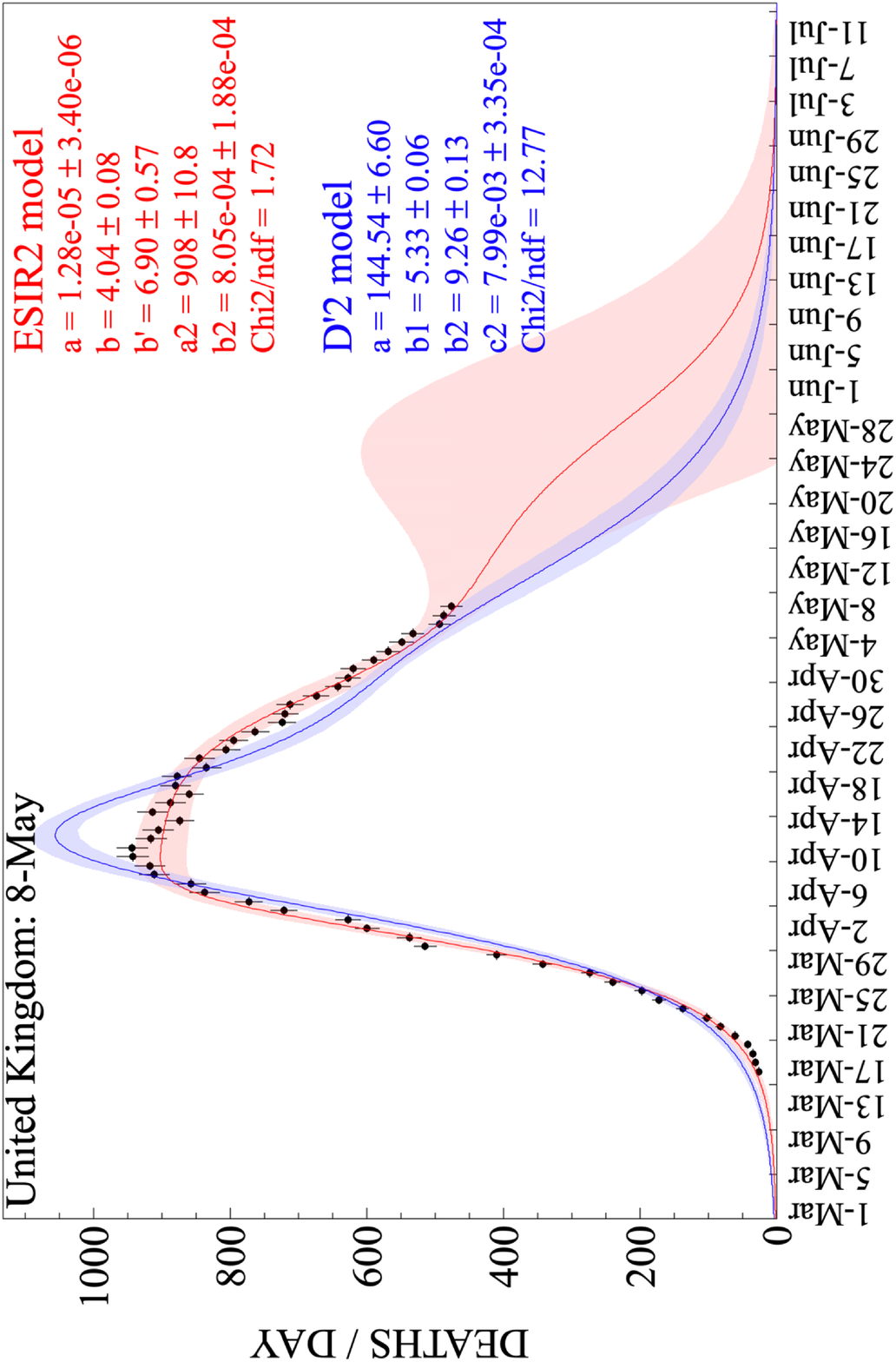}\\
\vspace{0.2cm}
\caption{Daily deaths fitted with the ESIR2 and $D^{\prime}_2$ models with a cut off of May 8 2020.}
\label{fig:total}       
\end{center}
\end{figure*}

Finally, Fig.~\ref{fig:total} shows the comparison between the ESIR2 and $D^{\prime}_2$ fits to real data for countries where {\small COVID-19} has widely spread: Belgium, USA, France, Germany, Iran, Italy, Spain and UK, USA.  Death data are taken from Refs. \cite{worldometer,spanish,french} and consider 7-day average smoothing to correct for anomalies in data collection such as the typical weekend staggering observed 
in various countries, where weekend data are counted at the beginning of the next week. 
Real error intervals are extracted from the correlation matrix.
As discussed in Section 2.3, the reduced $D^{\prime}_2$ model has been used with $a=a_2$ and $c=c_2$. 
Although arising from different assumptions, both models provide  similar
data descriptions and predictions, with slightly better values of $\chi^2$ per degree of freedom for the ESIR2 model.  
It is also interesting to note that the reduced ESIR2 model with five parameters yields similar results to the full ESIR2 model, 
with eight parameters.

As  data become available, daily predictions vary for both ESIR2 and the $D^{\prime}_2$ models. This is because the model
parameters are actually statistical averages over space-time of the 
properties of the complex system.  No model is able to predict changes over
time of these properties if the physical causes of these changes are
not included. The values of the model parameters are only
well defined when the disease spread is coming to an end and time
changes in the parameters have little influence.

\begin{figure*}[!ht] 
\begin{center}
\includegraphics[width=7.5cm,height=6cm,angle=-0]{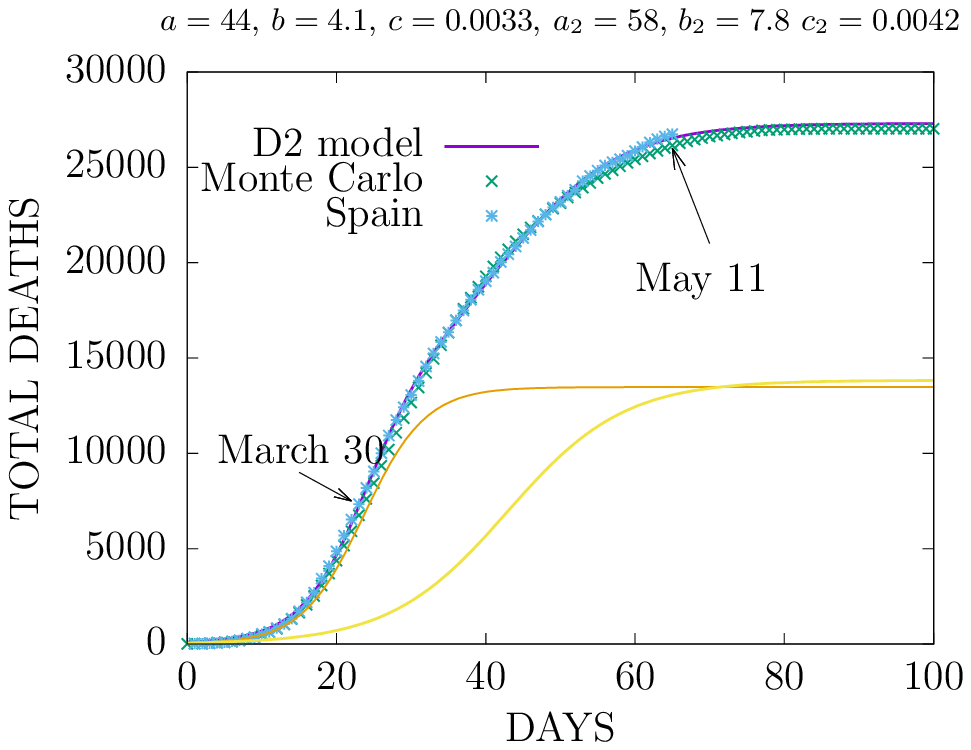}
\hspace{0.2cm} 
\includegraphics[width=7.5cm,height=6cm,angle=-0]{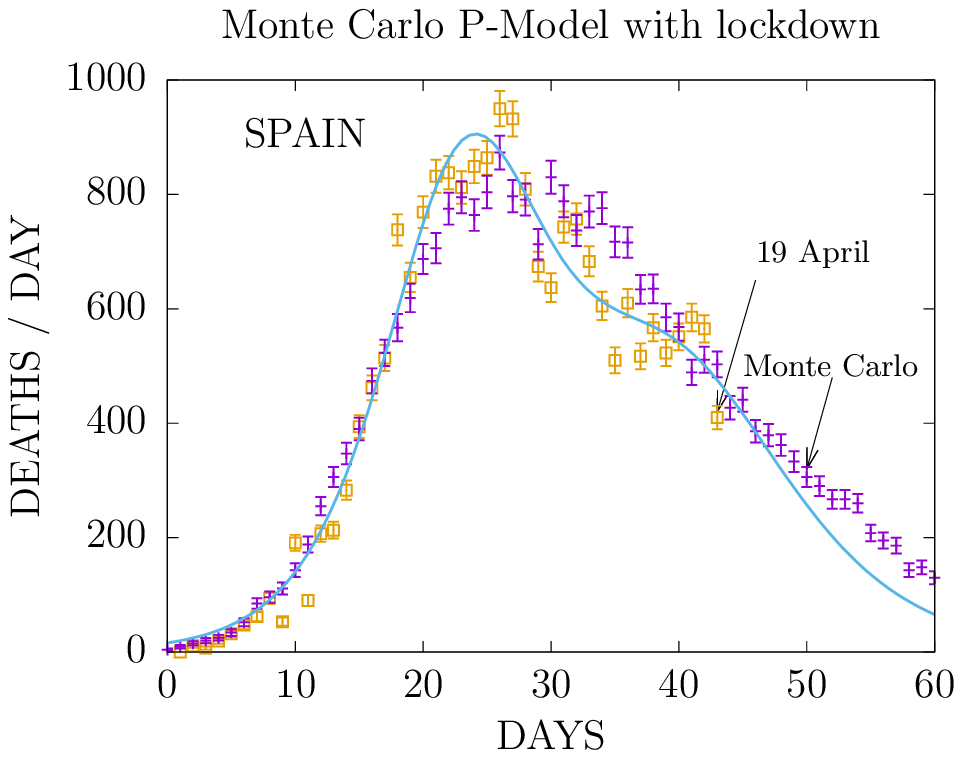}
\caption{Predictions of the MC and D models in Spain up to May 11.\label{mcp}
 }
\end{center}
\end{figure*}

\section{Discussion of Global Results}

More sophisticated calculations can be compared with ESIR2 and  $D^{\prime}_2$ predictions. 
In particular, Monte Carlo (MC) simulations have also been performed in this work for the Spanish case~\cite{amaro2}, 
which consist of a lattice of cells that can be in four different states: susceptible, infected, recovered or death. 
An infected cell can transmit the disease to any other susceptible cell within some random range $R$. 
The transmission mechanism follows principles of nuclear physics for the interaction of a particle with a target. 
Each infected particle interacts a number $n$ of times over the interaction region, according to its energy. 
The number of interactions is proportional to the interaction cross section $\sigma$ and to the target surface density 
$\rho$.  The discrete energy follows a Planck distribution law depending on the 'temperature' of the system.  
For any interaction, an infection probability is applied.  Finally, time-dependent recovery and death probabilities 
are also applied.  The resulting virus spread for different sets of parameters can be adjusted  
from {\small COVID-19} pandemic data. In addition, parameters can be made time 
dependent in order to investigate, for instance, the effect of an early lockdown or large mass gatherings at the rise of the 
pandemic. 

As shown in Fig.~\ref{mcp}, our MC simulations present similar results to the $D_2^{\prime}$ model, which 
validates the use of the simple D-model as a first-order approximation. More details on the MC simulation will be 
presented in a separate manuscript~\cite{amaro2}. Interestingly, MC simulations follow the data 
trend up to May 11 without any changes in the parameters for nearly two weeks.  
An app for Android devices, where the Monte Carlo Planck model has been implemented to visualize the simulation is 
available from Ref.~\cite{amaro3}.

In order to investigate the universality of the pandemic, it is interesting to compare all countries by plotting the $D$ model 
in terms of the variable $(t-t_0)/b$, where $t_0$ is the maximum of the daily curve given by 
$t_{max}= -b ~\ell n(c)$.  By shifting Eq.~\ref{D-model} by $t_{max}= -b ~\ell n(c)$ and dividing by $t_{max} = a/c$, the normalized  
$D$ function is given by, 
\begin{equation}
D_{norm}(t)=\frac{c~{\rm e}^{(t-t_{max})/b}}{1+c~{\rm e}^{(t-t_{max})/b}}. 
\end{equation}

\begin{figure*}[!ht]
\begin{center}
\includegraphics[width=11.cm,height=5cm,angle=-0]{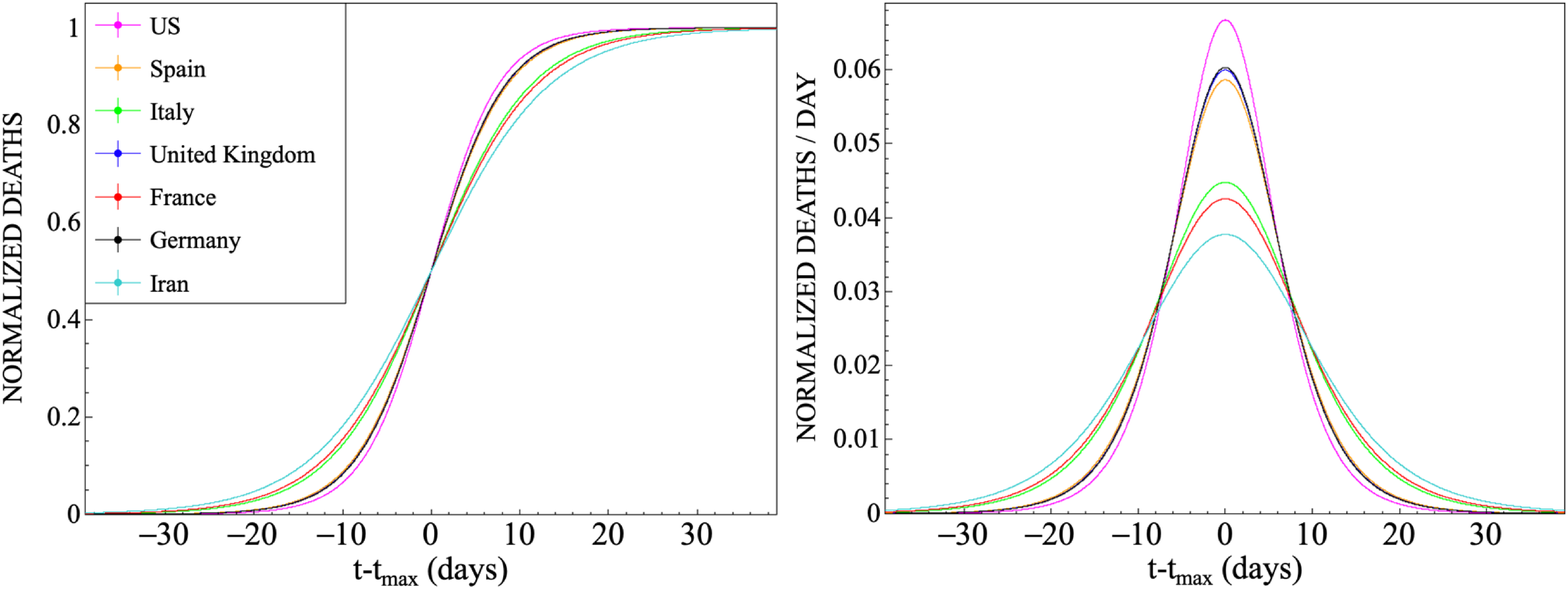}
\hspace{2mm}
\includegraphics[width=5.5cm,height=5cm,angle=-0]{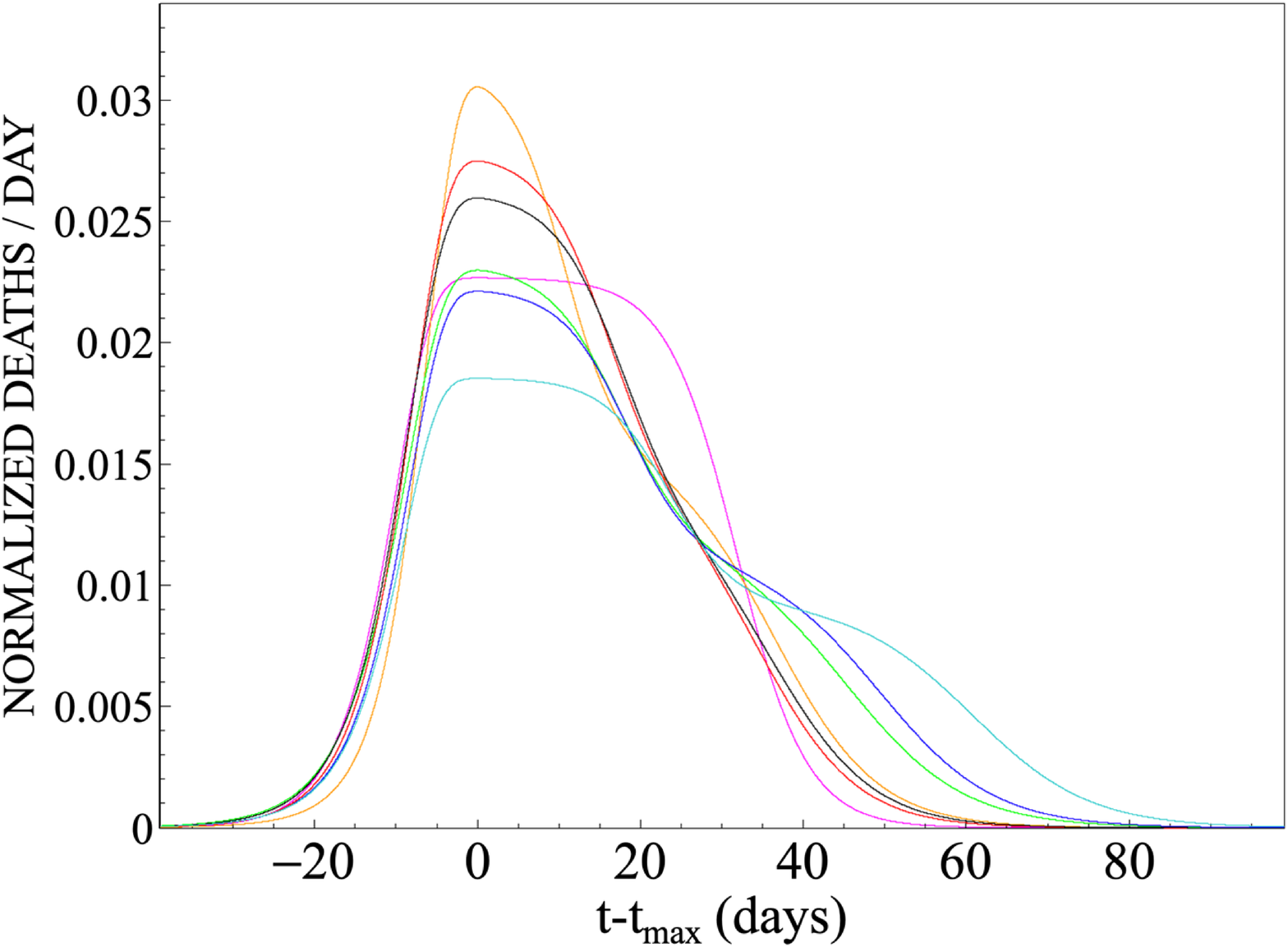}
\caption{Universality of the normalized $D$ (left), $D^{\prime}$ (middle) and  ESIR2 (right) models.}
\label{fig:universal}       
\end{center}
\end{figure*}

The left of Fig.~\ref{fig:universal} shows similar trends for the normalized D curves of different countries, which suggests a  
universal behavior of the {\small COVID-19} pandemic. Only Iran seems to slightly deviate from the global trend, which may indicate 
an early and more effective initial lockdown. A similar approach can be done for the daily data 
using the $D^{\prime}$ and ESIR2 models, as shown in the middle and right panels of Fig.~\ref{fig:universal}, respectively. 
Although different countries show similar trends, statistical fluctuations in the daily data do not result in a nice 
universal behavior as compared with $D_{norm}$. However, the $D^{\prime}$ and ESIR2 plots show that 
an effective lockdown is characterized by flatter and broader peaks, best characterized the Iranian case, whereas Spain and Germany  
present the sharper peaks.


%
%

\section{Final remarks}

%


The global models considered in this work present some differences with respect to other existing models. 
First, in this work we have tried to keep the models as simple as possible. 
This allows to use theoretical-inspired analytical expressions or semi-empirical formulae to perform the data analysis.
The use of semi-empirical expressions for describing physical phenomena is recurrent in physics.
One of the most famous is the semi-empirical mass formula from nuclear physics. Of course the free parameters
need to be fitted from known data, but this allowed to obtain predictions for unknown elements.

In our case we were inspired by the well known statistical SIR-kind models slightly modified to obtain analytical expressions that carry the leading time dependence.
  We have found that the $D$ and $D_2$ models allow a fast and efficient analysis of the pandemics in the initial and advanced stages.
Our results show that the time dependence of the pandemic parameters due to the lockdown can be effectively simulated
by the sum of two D-functions with different widths and heights and centered at different times. The distance between the
maxima of the two D-functions should be a measure of the time between the effective pandemic beginning and lockdown.

In the Spanish case this is about 20 days. Taking into account that lockdown started in March 14, this marks the pandemic starting time
as about February 22. Had the lockdown started on that date, the deaths would had been highly reduced.   
The smooth blending between the two peaks provides
a transition between the two statistical regimes  (or physical phases)  with and without lockdown.

The Monte Carlo simulation results are in agreement with our previous analysis with the $D$ and $D_2$ models.
The Monte Carlo generates events in a population of individuals in a lattice or grid of cells. We simulate the movement
of individuals outside of the cells and interactions with the susceptible individuals within a finite range. The randon events follow
statistical distributions based on the exponential laws of statistical mechanics for a system of interacting particles, driven by
macroscopic magnitudes as the temperature, and interaction probabilities between individuals, that can be related to interaction cross sections.

The Monte Carlo simulation spread the virus in space-time, and allows also space-time dependence on the parameters.
In this work we have made the simplest assumptions, only allowing for a lockdown effect by reducing the range of the interaction 
starting on a fixed day.
This simple modification allowed to reproduce nicely the Spanish death-per-day curve. The lockdown produces a relatively long broadening of 
the curve and a slow decay. Similar MC calculations can be performed in several countries to infer the devastating effect of a late 
lockdown as compared with early lockdown measures. The later is the case of South Africa and other countries, which have not reached 
the exponential growth.

The Death and extended SIR models are simple enough to provide 
fast estimations of pandemic evolution by fitting spatial-time average parameters, and 
present a good first-order approximation to understand secondary effects during the pandemic, 
such as lockdown and population migrations, which may help to control the disease. 
Similar models are available~\cite{ising,healthdata}, but challenges in epidemiological modeling 
remain~\cite{challenges,modeling,modeling2,harmonic}. This is a very complex system, 
which involves many degrees of freedom and millions of people, and even assuming consistent disease reporting - which is rarely the case -- 
there remains an important open question: Can any model predict the evolution of an epidemic from partial data?  Or similarly, 
Is it possible, at any given time and data, to measure the validity of an epidemic growth curve? 
We finally hope that we have added new insightful ideas with the Death, the extended SIR and Monte Carlo models, 
which can now be applied to any country which has followed the initial exponential pandemic growth.


\begin{acknowledgements}
The authors thank useful comments from Emmanuel Cl\'ement, Araceli Lopez-Martens, David Jenkins, 
Ramon Wyss, Liam Gaffney and Hans Fynbo. This work was supported by the Spanish Ministerio de Econom\'ia y Competitividad and European 
FEDER funds (grant FIS2017-85053-C2-1-P), Junta de Andaluc\'ia (grant FQM-225) and the South African National
Research Foundation (NRF) under Grant 93500.
\end{acknowledgements}

%
%



\end{document}